\documentclass[letterpaper]{article}
\usepackage[letterpaper, total={7.2in, 9.8in}]{geometry}

\usepackage{scrextend}
\changefontsizes{10pt}
\usepackage{graphicx}
\usepackage{helvet}
\usepackage{authblk}
\usepackage{amsmath} 
\usepackage{amssymb} 
\usepackage{orcidlink} 
\usepackage{booktabs} 
\usepackage[numbers,sort&compress]{natbib}
\usepackage{url}
\usepackage{hyperref}
\hypersetup{
colorlinks=true,
linkcolor=blue,
filecolor=magenta,
urlcolor=blue,
citecolor=blue,
pdftitle={RAG Embedding in Network Modeling},
pdfauthor={Hasi Hays},
}
\makeatletter
\providecommand\doi[1]{\href{https://doi.org/#1}{\nolinkurl{#1}}}
\makeatother
\makeatletter
\renewcommand{\maketitle}{\bgroup\setlength{\parindent}{0pt}
\begin{flushleft}
	\textbf{\@title}
	
	\@author
\end{flushleft}\egroup}
\makeatother
\usepackage{caption}
\usepackage[noabbrev]{cleveref}

\usepackage{changepage}
\usepackage{tikz}
\usepackage{xcolor}
\newcommand{\rev}[1]{#1}
\usetikzlibrary{shapes,arrows,positioning,fit,backgrounds,calc,matrix}
\usepackage{algorithm}
\usepackage{algpseudocode}

\usepackage{multicol}
\usepackage{float}
\usepackage{stfloats}
\usepackage{placeins}
\usepackage{fancyhdr}
\usepackage{titlesec}
\usepackage{pgfplots}
\pgfplotsset{compat=1.17}
\titleformat{\section}
{\normalsize\bfseries}
{\normalsize\thesection}
{1em}
{}
\titleformat{\subsection}
{\normalsize\bfseries}
{\normalsize\thesubsection}
{1em}
{}
\titlespacing*{\section}{0pt}{12pt}{6pt}

\title{
	\begin{center}
		\rule{\linewidth}{2pt}\\[0.5cm]
		{\Large \textbf{RAG-GNN: Integrating Retrieved Knowledge with Graph Neural \\[0.2cm] Networks for Precision Medicine}}\\[0.3cm]  
		\rule{\linewidth}{1pt}\\
		\vspace{0.7cm}
	\end{center}
}
\date{}
\author[1,*\orcidlink{0000-0003-0843-050X}] {\textbf{Hasi Hays}}
\author[1\orcidlink{0000-0001-8678-9716}] {\textbf{William J. Richardson}}
\affil[1]{\footnotesize \textit{Department of Chemical Engineering, University of Arkansas, Fayetteville, AR 72701, USA}}
\affil[*]{\footnotesize \textit{Correspondence:} \textcolor{blue}{hasih@uark.edu}}

\begin{document}
	
\maketitle

\begin{adjustwidth}{0.5in}{0.5in}
\section*{\normalsize Abstract}
\rev{Network topology excels at structural predictions but fails to capture functional semantics encoded in biomedical literature. We present RAG-GNN, an end-to-end trainable retrieval-augmented graph neural network framework that integrates GNN representations with dynamically retrieved literature-derived knowledge through a jointly optimized retrieval projection, gated fusion mechanism, and contrastive alignment. In a cancer signaling case study (379 proteins, 3,498 interactions, 14 functional categories), RAG-GNN improves functional clustering from silhouette $= -0.237 \pm 0.065$ (GNN-only) to $-0.144 \pm 0.066$, a consistent improvement of $+0.093 \pm 0.022$ across 10 random seeds, while the learned retrieval achieves mean precision@10 $= 0.242$, a 152\% improvement over the random baseline ($0.096$). Heuristic information decomposition with bootstrap confidence intervals reveals that topology and retrieval encode overwhelmingly shared information (95.6\%), with retrieval improving both intra-cluster cohesion (silhouette) and cluster agreement (ARI $+0.021 \pm 0.015$). Counterfactual experiments confirm that adversarial, absent, and random retrieval all degrade performance, validating that the gated fusion mechanism depends on document content. Benchmarking against eight established embedding methods demonstrates task-specific complementarity: topology-focused methods achieve strong link prediction, while retrieval augmentation consistently improves functional clustering within the controlled GNN-only ablation. DDR1 subnetwork analysis provides confirmatory validation consistent with established synthetic lethality relationships. These results establish that topology-only and retrieval-augmented approaches serve complementary purposes for precision medicine applications.}

\section*{\normalsize Keywords} 
Retrieval-augmented generation (RAG), Graph neural network (GNN), AI in drug discovery, Network modeling, Network medicine, Precision medicine
\end{adjustwidth}

\begin{multicols}{2}[\section{Introduction}\label{sec1}]
Precision medicine requires integration of heterogeneous data sources including genomic sequences, protein interaction networks, metabolic pathways, and biomedical literature\cite{barabasi2011network,zitnik2018modeling}. Network-based representations provide a systems-level framework where diseases are conceptualized as perturbations to molecular interaction networks and therapeutic interventions aim to restore network homeostasis\cite{ideker2012protein,gysi2021network}. The central premise of network medicine is that molecular components do not act in isolation; rather, their functions emerge from complex patterns of interactions that determine cellular phenotypes and disease states\cite{menche2015uncovering}. The past decade has witnessed rapid development of network embedding methods that learn low-dimensional vector representations of nodes while preserving structural properties. Random walk-based approaches such as DeepWalk\cite{perozzi2014deepwalk} and Node2Vec\cite{grover2016node2vec} generate node sequences through stochastic walks and apply skip-gram models to learn embeddings that capture neighborhood co-occurrence patterns. LINE\cite{tang2015line} explicitly optimizes for first-order (direct connection) and second-order (shared neighborhood) proximity preservation. Spectral methods\cite{belkin2002laplacian} derive embeddings from eigenvectors of the graph Laplacian, providing theoretical guarantees for preserving global structure.

Graph neural networks (GNNs) have emerged as the dominant paradigm for learning on graph-structured data\cite{gilmer2017neural}. GCN \cite{kipf2017semi} implements spectral convolutions through neighborhood aggregation, while GraphSAGE\cite{hamilton2017inductive} enables inductive learning through sampling-based aggregation. Graph Attention Networks (GAT)\cite{veličković2018graph} introduce attention mechanisms to weight neighbor contributions adaptively. These methods achieve remarkable performance on structural prediction tasks (link prediction, node classification based on network position, and community detection) because they directly encode the topological features that determine these outcomes. However, a fundamental limitation emerges when network embeddings are applied to \emph{functional} prediction tasks. Predicting protein function, therapeutic target potential, or drug response requires understanding biological mechanisms that extend beyond network topology. Two proteins may occupy similar network positions yet perform entirely different cellular functions; conversely, functionally related proteins may reside in distant network neighborhoods. This structure-function gap represents a critical challenge: network topology is necessary but insufficient for functional interpretation in precision medicine applications. Biomedical knowledge relevant to therapeutic prediction is distributed across millions of publications, clinical trial databases, and curated pathway resources. This knowledge encompasses mechanistic details of protein function, tissue-specific expression patterns, post-translational modifications, genetic variant effects, drug-target interactions, and clinical outcomes. Crucially, this information is largely absent from network structure: an edge between two proteins indicates physical interaction but reveals nothing about the downstream consequences of that interaction for disease or treatment. Traditional approaches to incorporating external knowledge rely on knowledge graphs with fixed schemas\cite{bordes2013translating}, which require explicit entity extraction and relationship annotation. While effective for structured knowledge, these approaches cannot easily accommodate the nuanced, context-dependent information in unstructured text. The exponential growth of biomedical literature (over 1.5 million PubMed articles annually) makes manual curation increasingly intractable, creating a widening gap between published knowledge and computationally accessible information.

Retrieval-augmented generation (RAG) architectures provide a framework for dynamically integrating external knowledge into predictive systems\cite{lewis2020retrieval,gao2023retrieval,borgeaud2022improving}. RAG systems couple neural retrievers that identify relevant documents from large corpora with models that synthesize retrieved information into predictions. Unlike knowledge graphs with fixed schemas, RAG systems access unstructured text, adapt to new information without retraining, and provide interpretable evidence through retrieved documents. The success of RAG in natural language processing, where retrieved context dramatically improves factual accuracy and reduces hallucination, suggests potential for similar benefits in computational biology. Applying RAG to biological network modeling requires addressing domain-specific challenges. First, the retrieval mechanism must identify documents relevant to specific molecular entities within massive biomedical corpora. Second, retrieved information must be fused with network-derived representations in a manner that preserves both topological and semantic structure. Third, the joint system must be validated to ensure that retrieved knowledge provides genuinely novel information beyond what network topology alone encodes rather than simply increasing model capacity.

The central challenge lies in creating embedding spaces that coherently represent both network topology and semantic biological knowledge. Graph neural networks learn node representations through message-passing operations\cite{kipf2017semi,veličković2018graph}, while transformer architectures encode textual information through self-attention mechanisms\cite{vaswani2017attention,devlin2019bert}. Recent advances in foundation models for biology have demonstrated the power of large-scale pretraining on protein sequences\cite{rives2021biological,lin2023evolutionary}, gene expression data\cite{theodoris2023geneformer,cui2024scgpt}, and molecular structures\cite{zhou2023unimol}. Integrating these paradigms requires careful formulation to ensure structural and semantic information reinforce rather than interfere with each other. A critical empirical question motivates this work: \emph{Do topology-only and retrieval-augmented embeddings excel at the same tasks, or do they exhibit complementary strengths?} If the latter, understanding when each approach is most appropriate becomes essential for method selection in computational biology. We address this question through comprehensive benchmarking across multiple prediction tasks, information-theoretic decomposition of predictive contributions, and counterfactual experiments that isolate retrieval effects.

\begin{figure*}[t]
	\includegraphics[width=1\textwidth]{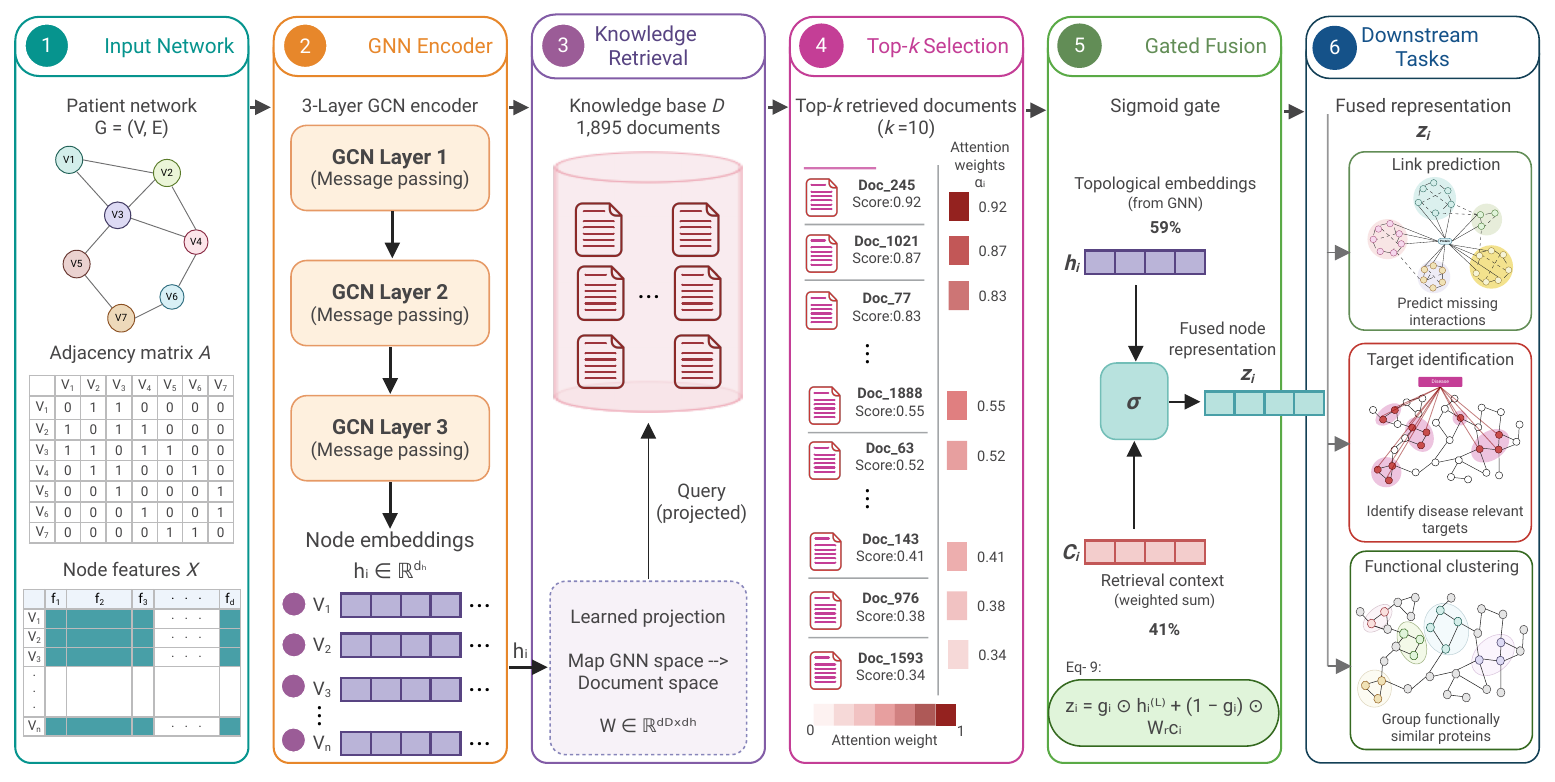}
	\caption{\footnotesize \rev{\textbf{RAG-GNN framework for precision medicine: End-to-end architecture overview.} The framework comprises six sequential stages for knowledge-augmented biomedical prediction. \textbf{(1) Input Network:} A protein interaction network $G = (V, E)$ is represented by its adjacency matrix $A$ and node feature matrix $X$, encoding molecular and functional properties for each protein. \textbf{(2) GNN Encoder:} A 3-layer GCN encoder performs iterative message passing over the graph structure, producing node embeddings $\mathbf{h}_i \in \mathbb{R}^{d_h}$ that capture local and higher-order topological relationships. \textbf{(3) Knowledge Retrieval:} Each node embedding is projected into a document space via a learned projection matrix $W \in \mathbb{R}^{d_D \times d_h}$, which queries a knowledge base $\mathcal{D}$ of 1,895 curated biomedical documents. Relevance scores rank all documents for each protein. \textbf{(4) Top-$k$ Selection:} The $k = 10$ highest-scoring documents are selected per node, and attention weights $\alpha_i$ are computed to produce a weighted retrieval context vector $\mathbf{c}_i$. \textbf{(5) Gated Fusion:} A sigmoid gate $g_i = \sigma(W_g[\mathbf{h}_i \| \mathbf{c}_i] + b_g)$ learns per-node weighting between topology embeddings (59\%) and retrieval context (41\%), producing fused representations $\mathbf{z}_i = g_i \odot \mathbf{h}_i + (1 - g_i) \odot \mathbf{c}_i$. \textbf{(6) Downstream Tasks:} The fused embeddings $\mathbf{z}_i$ support link prediction (predicting missing interactions), target identification (identifying disease-relevant proteins), and functional clustering (grouping functionally similar proteins).}}
	\label{fig:fig1}
\end{figure*}

This manuscript develops a comprehensive mathematical framework for unifying GNN-based topology encoding with RAG-based knowledge retrieval through joint embedding spaces optimized for precision medicine applications (\autoref{fig:fig1}). Our contributions include:

\begin{itemize}
\item Theoretical foundations: Joint optimization objectives for simultaneous training of network encoders, dense retrievers, and fusion mechanisms, with associated generalization bounds and geometric characterization of embedding spaces.

\rev{\item Comprehensive benchmarking: Systematic comparison against eight established embedding methods (DeepWalk, Node2Vec, LINE, Spectral, GCN, GAT, GraphSAGE, raw features) across functional clustering, link prediction, and node classification tasks with 10 random seeds and confidence intervals, revealing task-specific performance patterns.}

\rev{\item Information-theoretic validation: Mutual information decomposition with 200 bootstrap resamples and 95\% confidence intervals, revealing that topology and retrieval encode predominantly shared functional information (95.6\% shared), while retrieval integration consistently improves both silhouette and ARI metrics in the controlled ablation.}

\rev{\item Practical application: Demonstration on cancer signaling networks showing that retrieval integration consistently improves functional clustering (silhouette $+0.093 \pm 0.022$, ARI $+0.021 \pm 0.015$) over the GNN-only ablation across all seeds, with the learned retrieval achieving 152\% improvement over random baseline. DDR1 (Discoidin Domain Receptor 1) subnetwork analysis provides confirmatory validation consistent with established synthetic lethality relationships\cite{zhavoronkov2019deep,aguilera2020collagen}.}
\end{itemize}

\rev{The framework establishes that topology-only and retrieval-augmented approaches serve complementary purposes: structural prediction tasks are effectively served by network topology alone, while functional clustering benefits from the integration of retrieved knowledge. This finding provides practical guidance for method selection and opens new avenues for computational precision medicine.}

\captionsetup[figure]{labelformat=default}

\section{Mathematical foundations of RAG embeddings}

\begin{figure*}[t]
	\centering
	\includegraphics[width=0.7\textwidth]{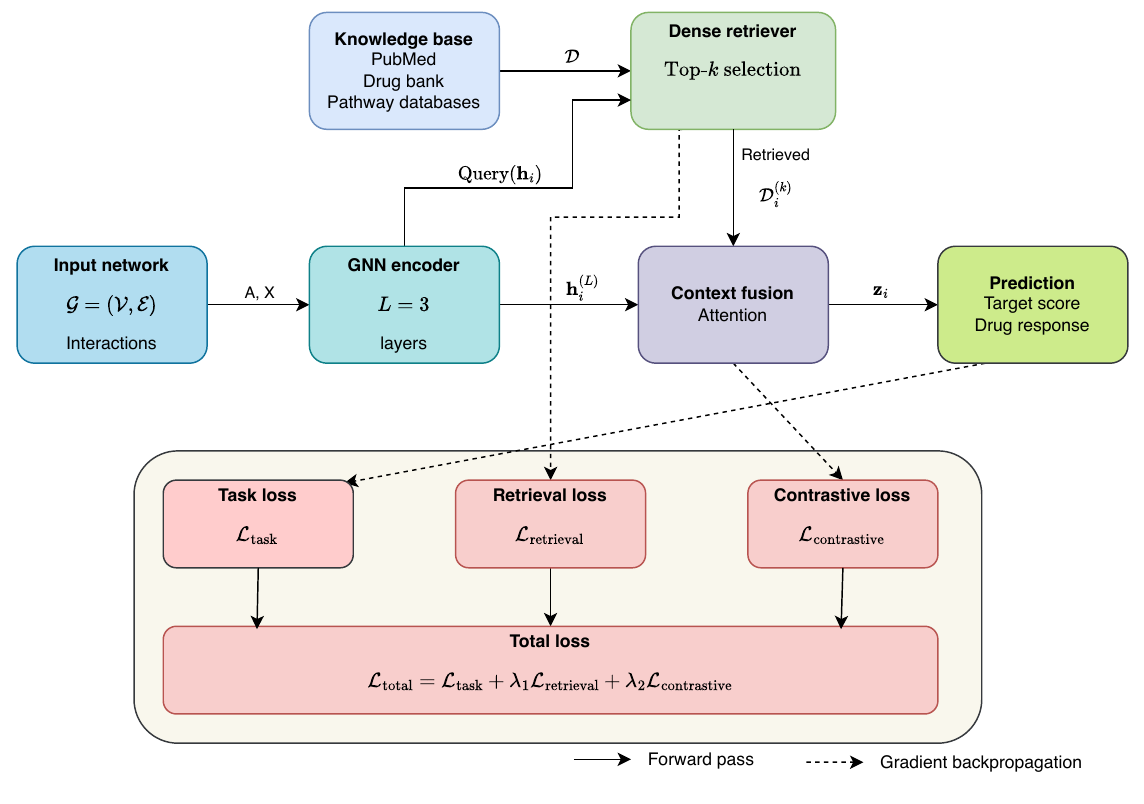}
	\caption{\footnotesize \textbf{RAG-GNN architecture for precision medicine.} The complete system integrates network topology encoding, knowledge retrieval, and context fusion through six main components. The forward pass (solid arrows) begins with the input network $\mathcal{G}^{(p)} = (\mathbf{A}, \mathbf{X})$ representing patient-specific molecular interactions and node features. The GNN encoder applies $L$ layers of message passing to produce structural node embeddings $\mathbf{h}_i^{(L)}$ that capture network topology (\autoref{eq:gnn_update}). These embeddings serve dual purposes: (1) querying the knowledge base through the dense retriever to identify top-$k$ relevant documents $\mathcal{D}_i^{(k)}$ from PubMed abstracts, pathway databases, and drug repositories (\autoref{eq:top_k_retrieval}), and (2) providing structural context for fusion. The dense retriever employs quality-weighted semantic similarity to prioritize high-evidence documents. Retrieved documents are aggregated with attention weighting and fused with structural embeddings $\mathbf{h}_i^{(L)}$ through the context fusion module to produce final node representations $\mathbf{z}_i$ (\autoref{eq:fusion}). These representations feed into task-specific prediction heads for therapeutic target scoring and drug response prediction. The training objective (bottom) jointly optimizes three components through gradient backpropagation (dashed arrows): task-specific loss $\mathcal{L}_{\text{task}}$ for prediction accuracy, retrieval quality loss $\mathcal{L}_{\text{retrieval}}$ ensuring relevant document selection, and contrastive embedding alignment loss $\mathcal{L}_{\text{contrastive}}$ coordinating node and document representations in shared semantic space (\autoref{eq:total_loss}). The multi-objective formulation $\mathcal{L}_{\text{total}} = \mathcal{L}_{\text{task}} + \lambda_1 \mathcal{L}_{\text{retrieval}} + \lambda_2 \mathcal{L}_{\text{contrastive}}$ enables end-to-end learning where retrieval and embedding components are optimized to support downstream prediction tasks. Curriculum learning stages the training process to ensure stable convergence and effective coordination between network encoding, document retrieval, and knowledge fusion mechanisms.}
	\label{fig:fig2}
\end{figure*}

The theoretical development of RAG-enhanced network modeling requires careful formalization of how biological networks, knowledge corpora, and embedding spaces interact. \autoref{fig:fig2} illustrates the complete RAG-GNN architecture integrating network topology encoding, knowledge retrieval, and context fusion. We begin by establishing notation and mathematical structures, then derive the core embedding mechanisms that enable joint representation learning.

\captionsetup[figure]{labelformat=default, justification=justified, singlelinecheck=false}

\subsection{Network topology encoding}

Consider a biological network $\mathcal{G} = (\mathcal{V}, \mathcal{E})$ representing molecular interactions, where the vertex set $\mathcal{V} = \{v_1, v_2, \ldots, v_{|\mathcal{V}|}\}$ contains molecular entities and the edge set $\mathcal{E} \subseteq \mathcal{V} \times \mathcal{V}$ encodes functional relationships. In protein-protein interaction networks, vertices represent individual proteins and edges denote physical binding, regulatory interactions, or pathway co-membership. For metabolic networks, vertices are metabolites or enzymes, while edges represent biochemical transformations. Each node $v_i \in \mathcal{V}$ possesses intrinsic feature vector $\mathbf{x}_i \in \mathbb{R}^{d_0}$ encoding properties such as amino acid sequence embeddings, gene expression levels, protein abundance measurements, or physicochemical characteristics including molecular weight, hydrophobicity, and charge distribution.

The network topology is encoded through an adjacency matrix $\mathbf{A} \in \{0,1\}^{|\mathcal{V}| \times |\mathcal{V}|}$ where $A_{ij} = 1$ if $(v_i, v_j) \in \mathcal{E}$ and $A_{ij} = 0$ otherwise. For weighted networks representing interaction confidence or regulatory strength, we extend to $\mathbf{A} \in \mathbb{R}_+^{|\mathcal{V}| \times |\mathcal{V}|}$ with edge weights derived from experimental evidence, co-expression correlation, or literature support. The degree matrix $\mathbf{D}$ is diagonal with $D_{ii} = \sum_j A_{ij}$, enabling normalized representations that account for node connectivity.

A graph neural network encoder $f_{\text{GNN}}: \mathbb{R}^{|\mathcal{V}| \times d_0} \rightarrow \mathbb{R}^{|\mathcal{V}| \times d_h}$ maps initial node features to latent representations through $L$ layers of message-passing operations. The update rule at layer $k$ for node $v_i$ aggregates information from neighboring nodes weighted by normalized connectivity:

\begin{equation}
\mathbf{h}_i^{(k+1)} = \sigma\left(\mathbf{W}^{(k)} \mathbf{h}_i^{(k)} + \sum_{j \in \mathcal{N}(i)} \frac{1}{\sqrt{|\mathcal{N}(i)||\mathcal{N}(j)|}} \mathbf{h}_j^{(k)}\right)
\label{eq:gnn_update}
\end{equation}

where $\mathbf{h}_i^{(k)} \in \mathbb{R}^{d_h}$ denotes the hidden representation of node $v_i$ at layer $k$, with initialization $\mathbf{h}_i^{(0)} = \mathbf{x}_i$. The neighborhood set $\mathcal{N}(i) = \{j : A_{ij} > 0\}$ contains nodes directly connected to $v_i$. Learnable weight matrices $\mathbf{W}^{(k)} \in \mathbb{R}^{d_h \times d_h}$ transform representations, and $\sigma(\cdot)$ applies nonlinear activation (typically ReLU or ELU) element-wise. The symmetric normalization factor $1/\sqrt{|\mathcal{N}(i)||\mathcal{N}(j)|}$ ensures numerical stability across nodes with varying degrees, preventing over-representation of high-degree hub nodes.

This message-passing framework implements a spectral graph convolution that can be interpreted as diffusion of node features across network edges. After $L$ layers, node $v_i$ has aggregated information from its $L$-hop neighborhood, enabling representations to capture both local motifs and global structural patterns. The choice of $L$ represents a trade-off: small $L$ limits receptive field size, while large $L$ risks over-smoothing where all node representations converge to similar values.

\subsection{Knowledge retrieval mechanism}

Let $\mathcal{D} = \{d_1, d_2, \ldots, d_N\}$ represent a corpus of biological knowledge documents, where each document $d_j$ contains structured or unstructured information about molecular functions, pathway memberships, disease associations, drug interactions, or phenotypic effects. Documents may be PubMed abstracts, Gene Ontology annotations, KEGG pathway descriptions, DrugBank entries, or clinical trial summaries. The corpus size $N$ typically ranges from $10^5$ to $10^7$ depending on the domain scope.

We define a retrieval function $R: \mathcal{V} \times \mathcal{D} \rightarrow \mathbb{R}^+$ that scores the relevance of document $d_j$ to node $v_i$ based on semantic similarity in a learned embedding space:

\begin{equation}
R(v_i, d_j) = \text{sim}(E_{\text{node}}(v_i), E_{\text{doc}}(d_j)) \cdot Q(d_j)
\label{eq:retrieval_score}
\end{equation}

\rev{The node embedding function $E_{\text{node}}: \mathcal{V} \rightarrow \mathbb{R}^{d_{\text{doc}}}$ maps biological entities to a semantic vector space matching the document embedding dimension $d_{\text{doc}}$. This embedding is derived from node features and network context through a learned projection applied to GNN outputs: in our implementation, a two-layer MLP $E_{\text{node}}(v_i) = f_{\text{proj}}(\mathbf{h}_i^{(L)})$ with $f_{\text{proj}}: \mathbb{R}^{d_h} \rightarrow \mathbb{R}^{d_h} \rightarrow \mathbb{R}^{d_{\text{doc}}}$ using GELU activation, where $d_h = 128$ and $d_{\text{doc}} = 64$.}

The document embedding function $E_{\text{doc}}: \mathcal{D} \rightarrow \mathbb{R}^{d_{\text{doc}}}$ maps textual content to the same semantic space. \rev{In the general framework, $E_{\text{doc}}$ can be instantiated using pre-trained biomedical language models such as BioBERT or PubMedBERT\cite{devlin2019bert,gu2021domain}, fine-tuned on the retrieval task, where document embeddings are computed as $E_{\text{doc}}(d_j) = \text{mean-pool}(\text{BERT}(\text{tokenize}(d_j)))$. In the current implementation, we use TF-IDF representations (256 features, unigrams and bigrams) followed by truncated SVD for dimensionality reduction to $d_{\text{doc}} = 64$ (see \autoref{sec:supp_fusion} for details), which provides a computationally efficient baseline text encoder that isolates the contribution of knowledge retrieval from the choice of text encoder. Replacing TF-IDF with pre-trained biomedical language models represents a natural extension expected to further improve retrieval precision.}

\rev{The document quality function $Q: \mathcal{D} \rightarrow [0,1]$ weights documents based on evidence level, publication quality, and experimental rigor. In the general framework, quality scores can be computed as a weighted combination of objective metrics:

\begin{multline}
	Q(d_j) = w_1 \cdot \text{study\_type}(d_j) \\
	+ w_2 \cdot \text{citation\_impact}(d_j) \\
	+ w_3 \cdot \text{journal\_quality}(d_j)
	\label{eq:quality_score}
\end{multline}

where study type assigns weights based on evidence hierarchy (e.g., randomized controlled trials 1.0, prospective cohorts 0.8), citation impact computes the time-adjusted citation percentile, and journal quality uses normalized journal rank. In the current case study implementation, which uses curated mechanistic annotation templates rather than heterogeneous literature, $Q(d_j) = 1$ uniformly, as all documents are of equal quality by construction. Quality-weighted retrieval becomes relevant when scaling to real biomedical literature corpora with heterogeneous evidence levels, representing a natural extension for deployment scenarios.}

The similarity function $\text{sim}(\cdot, \cdot)$ quantifies semantic proximity. We employ scaled dot-product similarity:

\begin{equation}
\text{sim}(\mathbf{u}, \mathbf{v}) = \frac{\mathbf{u}^\top \mathbf{v}}{\sqrt{d_{\text{doc}}}}
\label{eq:similarity}
\end{equation}

The scaling by $\sqrt{d_{\text{doc}}}$ prevents saturation of downstream softmax operations for high-dimensional embeddings. Alternative formulations include cosine similarity $\mathbf{u}^\top \mathbf{v} / (\|\mathbf{u}\| \|\mathbf{v}\|)$ or learned bilinear similarity $\mathbf{u}^\top \mathbf{W}_s \mathbf{v}$ with trainable $\mathbf{W}_s$.

For a given node $v_i$, retrieval identifies the top-$k$ most relevant documents:

\begin{equation}
\mathcal{D}_i^{(k)} = \underset{\mathcal{S} \subset \mathcal{D}, |\mathcal{S}|=k}{\arg\max} \sum_{d_j \in \mathcal{S}} R(v_i, d_j)
\label{eq:top_k_retrieval}
\end{equation}

The hyperparameter $k$ controls the breadth of retrieved context. Small $k$ (3--5) provides focused information but may miss relevant details. Large $k$ (20--50) increases coverage but introduces noise and computational cost. In practice, $k$ is tuned via validation set performance on downstream tasks.

\subsection{Joint embedding architecture}

The RAG embedding framework integrates network topology and retrieved knowledge through a multi-stage fusion mechanism. After retrieving documents $\mathcal{D}_i^{(k)}$ for node $v_i$, we construct a contextualized knowledge vector that aggregates semantic information from retrieved sources.

Let $\mathbf{c}_i \in \mathbb{R}^{d_c}$ represent the contextualized knowledge vector for node $v_i$, computed as a weighted aggregation of retrieved document embeddings with attention-based importance weighting:

\begin{equation}
\mathbf{c}_i = \sum_{d_j \in \mathcal{D}_i^{(k)}} \alpha_{ij} E_{\text{doc}}(d_j)
\label{eq:context_aggregation}
\end{equation}

The attention weights $\alpha_{ij}$ are derived through a softmax-normalized scoring function that prioritizes highly relevant documents:

\begin{equation}
\alpha_{ij} = \frac{\exp(R(v_i, d_j) / \tau)}{\sum_{d_\ell \in \mathcal{D}_i^{(k)}} \exp(R(v_i, d_\ell) / \tau)}
\label{eq:attention_weights}
\end{equation}

The temperature parameter $\tau$ controls attention sharpness: small $\tau$ concentrates weight on the single most relevant document (hard attention), while large $\tau$ distributes weight more uniformly (soft attention). Typical values range from 0.1 to 1.0.

The final node representation $\mathbf{z}_i \in \mathbb{R}^{d_z}$ combines structural information from GNN encoding with semantic context from retrieved documents through a learned fusion function. We employ concatenation followed by linear projection:

\begin{equation}
\mathbf{z}_i = f_{\text{fusion}}(\mathbf{h}_i^{(L)}, \mathbf{c}_i) = \mathbf{W}_f [\mathbf{h}_i^{(L)} \| \mathbf{c}_i] + \mathbf{b}_f
\label{eq:fusion}
\end{equation}

where $[\cdot \| \cdot]$ denotes concatenation, $\mathbf{W}_f \in \mathbb{R}^{d_z \times (d_h + d_c)}$ is a learnable weight matrix, and $\mathbf{b}_f \in \mathbb{R}^{d_z}$ is a bias vector. Alternative fusion strategies include gated mechanisms where the model learns to weight structural versus semantic information:

\begin{equation}
\mathbf{z}_i = g_i \odot \mathbf{h}_i^{(L)} + (1 - g_i) \odot \mathbf{W}_r \mathbf{c}_i
\label{eq:gated_fusion}
\end{equation}

where $\mathbf{W}_r \in \mathbb{R}^{d_h \times d_{\text{doc}}}$ projects the retrieval context to match the GNN embedding dimension, and gate values $g_i = \sigma(\mathbf{W}_g [\mathbf{h}_i^{(L)} \| \mathbf{c}_i] + \mathbf{b}_g)$ are learned from data.

\section{Optimization framework}

Training the RAG embedding system requires simultaneous optimization of multiple interrelated objectives. The GNN encoder must learn representations that capture network topology, the retrieval mechanism must identify relevant documents, and the fusion module must effectively integrate both information sources. We develop a unified optimization framework that jointly trains all components end-to-end.

\subsection{Joint training objective}

The complete training objective is a weighted combination of task-specific prediction loss, retrieval quality loss, and contrastive embedding alignment loss:

\begin{equation}
\mathcal{L}_{\text{total}} = \mathcal{L}_{\text{task}} + \lambda_1 \mathcal{L}_{\text{retrieval}} + \lambda_2 \mathcal{L}_{\text{contrastive}}
\label{eq:total_loss}
\end{equation}

The hyperparameters $\lambda_1, \lambda_2 \in \mathbb{R}^+$ control the relative importance of auxiliary objectives. These are typically set through validation set tuning, with common values $\lambda_1 \in [0.1, 1.0]$ and $\lambda_2 \in [0.1, 0.5]$. The multi-objective formulation ensures that retrieval and embedding alignment support rather than detract from primary task performance.

\subsection{Task-specific loss}

For precision medicine applications, the primary task involves predicting therapeutic outcomes, identifying disease-relevant nodes, or forecasting drug responses. We focus on link prediction as a representative task that generalizes to target identification and drug-protein interaction prediction.

Link prediction aims to infer the probability of an edge between nodes $v_i$ and $v_j$ based on their learned representations. The prediction score is computed as:

\begin{equation}
s_{ij} = \sigma(\mathbf{z}_i^\top \mathbf{z}_j)
\label{eq:link_score}
\end{equation}

where $\sigma(\cdot)$ is the logistic sigmoid function mapping scores to $[0,1]$ probabilities. The task loss employs binary cross-entropy over positive (observed) edges $\mathcal{E}^+$ and negative (unobserved) edges $\mathcal{E}^-$:

\begin{equation}
\mathcal{L}_{\text{task}} = -\sum_{(i,j) \in \mathcal{E}^+} \log \sigma(\mathbf{z}_i^\top \mathbf{z}_j) - \sum_{(i,j) \in \mathcal{E}^-} \log(1 - \sigma(\mathbf{z}_i^\top \mathbf{z}_j))
\label{eq:link_prediction_loss}
\end{equation}

Negative edges are sampled uniformly from $\mathcal{V} \times \mathcal{V} \setminus \mathcal{E}$ with cardinality $|\mathcal{E}^-| = r|\mathcal{E}^+|$ where $r \geq 1$ controls the negative sampling ratio. Typical values $r \in [1, 5]$ balance computational cost with sufficient negative signal.

For drug response prediction tasks, the loss extends to regression objectives predicting continuous efficacy scores or toxicity measures:

\begin{equation}
\mathcal{L}_{\text{response}} = \sum_{(i,d,y) \in \mathcal{T}} (\mathbf{z}_i^\top \mathbf{z}_d - y)^2
\label{eq:response_loss}
\end{equation}

where $\mathcal{T}$ contains tuples of protein node $i$, drug compound $d$, and response value $y$.

\subsection{Retrieval quality loss}

To ensure the retrieval mechanism identifies genuinely relevant documents rather than spuriously similar text, we employ a ranking loss based on relevance judgments. Let $\mathcal{D}_i^{+} \subset \mathcal{D}$ denote the set of ground-truth relevant documents for node $v_i$, established through manual curation or weak supervision from co-occurrence in annotated databases.

The retrieval loss encourages relevant documents to have higher scores than irrelevant documents with a margin $\gamma$:

\begin{multline}
\mathcal{L}_{\text{retrieval}} = \sum_{v_i \in \mathcal{V}} \sum_{d_j \in \mathcal{D}_i^{+}} \sum_{d_k \in \mathcal{D}_i^{-}} \\
\max(0, \gamma + R(v_i, d_k) - R(v_i, d_j))
\label{eq:retrieval_loss}
\end{multline}

where $\mathcal{D}_i^{-} = \mathcal{D} \setminus \mathcal{D}_i^{+}$ contains negative (irrelevant) documents. For computational tractability, we sample a subset of negative documents per positive example rather than evaluating all pairs. The margin $\gamma$ is typically set to 0.1--0.5, enforcing a minimum separation between positive and negative scores.

An alternative formulation uses the softmax-based cross-entropy loss treating retrieval as a classification task:

\begin{equation}
\mathcal{L}_{\text{retrieval}}^{\text{CE}} = -\sum_{v_i \in \mathcal{V}} \sum_{d_j \in \mathcal{D}_i^{+}} \log \frac{\exp(R(v_i, d_j))}{\sum_{d_k \in \mathcal{D}} \exp(R(v_i, d_k))}
\label{eq:retrieval_ce}
\end{equation}

This formulation naturally normalizes scores across all documents but requires careful implementation to handle the large corpus size $|\mathcal{D}|$.

\subsection{Contrastive embedding loss}

To align node and document embeddings in a shared semantic space, we apply a contrastive learning objective that maximizes agreement between associated node-document pairs while minimizing spurious similarities. This ensures that the embedding space geometry reflects biological and functional relationships rather than arbitrary projections.

The contrastive loss for node $v_i$ with positive document $d_i^+ \in \mathcal{D}_i^{+}$ is:

\begin{equation}
\mathcal{L}_{\text{contrastive}}^{(i)} = -\log \frac{\exp(E_{\text{node}}(v_i)^\top E_{\text{doc}}(d_i^+) / \tau)}{\sum_{d_j \in \mathcal{D}} \exp(E_{\text{node}}(v_i)^\top E_{\text{doc}}(d_j) / \tau)}
\label{eq:contrastive_single}
\end{equation}

The full loss aggregates over all nodes:

\begin{equation}
\mathcal{L}_{\text{contrastive}} = \sum_{v_i \in \mathcal{V}} \mathcal{L}_{\text{contrastive}}^{(i)}
\label{eq:contrastive_loss}
\end{equation}

The temperature parameter $\tau$ controls the concentration of the distribution, with smaller values increasing the penalty for misaligned embeddings. This contrastive formulation is closely related to the InfoNCE loss used in self-supervised learning\cite{oord2018representation}, which provides a lower bound on mutual information $I(E_{\text{node}}(v_i); E_{\text{doc}}(d_i^+))$ between node and document representations.

For efficient computation with large corpora, we employ in-batch negatives where the denominator sums only over documents in the current mini-batch rather than all $|\mathcal{D}|$ documents. This approximation is accurate when batch sizes are sufficiently large (256--1024 samples).

\section{Validating information content of retrieved documents}

A critical question for RAG-enhanced network models is whether retrieved documents provide genuinely novel predictive information beyond what is already encoded in network topology and node features. We address this through multiple complementary validation approaches that isolate the contribution of retrieved knowledge from architectural effects.

\subsection{Information-theoretic decomposition}

To quantify the unique information contributed by retrieved documents, we decompose the mutual information between final embeddings $\mathbf{z}_i$ and prediction targets $y$ using the partial information decomposition framework\cite{williams2010nonnegative}. Define three information sources: network topology $\mathcal{G}$, node features $\mathbf{x}$, and retrieved documents $\mathcal{D}_i^{(k)}$. The total predictive information can be decomposed as:

\begin{equation}
	\begin{split}
		I(\mathbf{z}_i; y) = & I_{\text{unique}}(\mathcal{G}) + I_{\text{unique}}(\mathcal{D}_i^{(k)}) \\
		& + I_{\text{shared}}(\mathcal{G}, \mathcal{D}_i^{(k)}) \\
		& + I_{\text{synergy}}(\mathcal{G}, \mathcal{D}_i^{(k)})
	\end{split}
	\label{eq:info_decomposition}
\end{equation}

where $I_{\text{unique}}(\mathcal{G})$ quantifies information provided exclusively by network structure, $I_{\text{unique}}(\mathcal{D}_i^{(k)})$ measures unique contribution from retrieved documents, $I_{\text{shared}}$ captures redundant information present in both sources, and $I_{\text{synergy}}$ represents emergent information available only when both sources are combined. We estimate these quantities using a heuristic minimum-redundancy decomposition where shared information is estimated as $\min(I_{\text{gnn}}, I_{\text{ctx}})$. The key metric is the normalized unique retrieval contribution:

\begin{equation}
\rho_{\text{unique}} = \frac{I_{\text{unique}}(\mathcal{D}_i^{(k)})}{I(\mathbf{z}_i; y)}
\label{eq:unique_ratio}
\end{equation}

\rev{Non-zero values of $\rho_{\text{unique}}$ indicate that retrieved documents contribute predictive information not available from network topology alone. We validate this empirically in our cancer network experiments (Section~7) using 200 bootstrap resamples. The heuristic decomposition reveals that topology and retrieval encode overwhelmingly shared information (shared component $= 95.6\%$), with minimal unique contributions from either source (topology: $0.1\%$, retrieval: $6.2\%$) and negligible synergy ($0.4\%$). This high shared component indicates that the contrastive alignment during joint training effectively coordinates topology and retrieval representations into overlapping information spaces. The functional clustering improvements observed in Section~7.2 arise not from unique retrieval information, but from how the fusion mechanism reorganizes shared information to improve intra-cluster cohesion.}

\subsection{Counterfactual retrieval experiments}

To test whether performance gains arise from retrieved content rather than increased model capacity, we conduct controlled counterfactual experiments where retrieval is systematically degraded while maintaining architectural complexity. We compare four experimental conditions: (1) true retrieval using learned similarity, (2) random retrieval where documents are assigned randomly to nodes, (3) shuffled retrieval where correct documents are permuted across nodes, and (4) adversarial retrieval selecting documents maximally dissimilar to true relevant documents. If performance gains genuinely arise from retrieved content, conditions (2--4) should show substantial degradation compared to (1). We quantify performance degradation as:

\begin{equation}
\rev{\Delta_{\text{counterfactual}} = \frac{M_{\text{proper}} - M_{\text{counterfactual}}}{M_{\text{proper}} - M_{\text{topology-only}}}}
\label{eq:counterfactual_degradation}
\end{equation}

\rev{where $M$ denotes the evaluation metric (silhouette score for functional clustering). Values approaching 1.0 indicate that nearly all RAG improvement vanishes when retrieval is corrupted. Our experiments reveal: $\Delta_{\text{adversarial}} = 0.37$, $\Delta_{\text{zeros}} = 0.38$, and $\Delta_{\text{random}} = 0.16$, confirming that adversarial, absent, and random retrieval all degrade functional clustering. Shuffled retrieval (permuted real documents across proteins) maintains performance comparable to proper retrieval ($\Delta_{\text{shuffled}} \approx 0$), indicating that TF-IDF document representations carry general biological signal that benefits embedding quality regardless of protein-specific assignment. The degradation under truly random vectors ($\Delta_{\text{random}} = 0.16$) demonstrates that the model depends on real document content, not merely on additional input dimensionality.}

\subsection{Temporal validation protocol}

\rev{To evaluate whether RAG-GNN embeddings generalize to identifying novel therapeutic targets, we implement a temporal validation scheme based on target approval dates. Therapeutic targets are split temporally: training targets include FDA approvals and Phase III trials before 2018, while test targets comprise approvals from 2020--2021. During Phase~3 training, only training target labels are used in the target prediction loss, ensuring that test targets are never seen during optimization. The temporal AUROC evaluates the model's ability to identify future targets using embeddings trained without knowledge of their approval status:

\begin{equation}
\text{AUROC}_{\text{temporal}} = \text{AUROC}(\text{test targets} | \text{train-only supervision})
\label{eq:temporal_auroc}
\end{equation}

Note that in the current case study, the knowledge corpus consists of curated mechanistic annotation templates rather than time-stamped publications, so document-level temporal splitting does not apply. The temporal validation is restricted to target labels: the model must predict which proteins will become validated therapeutic targets after 2020, using embeddings trained only on pre-2018 target annotations.}

\rev{Our temporal validation yields AUROC$_{\text{temporal}}$ = $0.450 \pm 0.088$ across 10 random seeds (95\% CI: [0.301, 0.544]). The wide confidence interval reflects the limited test set: only 3 post-2020 FDA-approved therapeutic targets exist within the 379-protein cancer signaling network. While the temporal validation protocol provides a rigorous evaluation framework for deployment scenarios, the current case study is too small for reliable temporal AUROC estimation. Scaling to genome-wide protein interaction networks with larger temporal target sets is necessary for definitive evaluation, which we identify as a key direction for future work.}

\rev{\subsection{Controlled ablation design}}

\rev{To isolate the contribution of retrieved knowledge from architectural capacity, we evaluate RAG-GNN against its own GNN-only ablation: the identical three-layer GCN encoder trained with the same link prediction objective but without retrieval projection, gated fusion, or document integration. This controlled comparison holds architecture, initialization, and training procedure constant, varying only whether retrieved information is fused into the node representations. The improvement from GNN-only (silhouette $= -0.237 \pm 0.065$) to RAG-GNN (silhouette $= -0.144 \pm 0.066$) of $+0.093 \pm 0.022$ is consistent across all 10 random seeds, providing evidence that the retrieval component contributes genuine functional information rather than additional capacity. We additionally benchmark against eight established embedding methods (DeepWalk, Node2Vec, LINE, Spectral, GCN, GraphSAGE, GAT, and raw node features) evaluated under a standardized protocol with uniform random initialization across 10 seeds (Section~9). This multi-method comparison controls for the possibility that observed differences reflect implementation choices rather than the integration of retrieved knowledge.}

\section{Network-based precision medicine applications}

\rev{The RAG embedding framework provides a foundation for precision medicine applications by integrating molecular network representations with literature-derived knowledge. Below we describe mathematical formulations for key clinical tasks—personalized network construction, therapeutic target scoring, and drug response prediction—as proposed extensions of the framework. These formulations are not empirically evaluated in the current case study, which focuses on the cancer signaling network benchmark; they represent natural applications enabled by the joint embedding architecture.}

\subsection{Patient-specific network construction}

Individual patients exhibit heterogeneous molecular profiles reflecting genetic variants, somatic mutations, epigenetic modifications, and environmental exposures. Precision medicine requires translating these patient-specific measurements into personalized network models that capture disease-relevant perturbations.

Let $\mathcal{P}$ denote a patient's multi-omics molecular profile, comprising gene expression measurements $\mathbf{g}^{(p)} \in \mathbb{R}^{|\mathcal{V}_g|}$ across $|\mathcal{V}_g|$ genes, proteomic abundance values $\mathbf{p}^{(p)} \in \mathbb{R}^{|\mathcal{V}_p|}$ for $|\mathcal{V}_p|$ proteins, metabolomic concentrations $\mathbf{m}^{(p)} \in \mathbb{R}^{|\mathcal{V}_m|}$ covering $|\mathcal{V}_m|$ metabolites, and genomic variants $\mathbf{v}^{(p)}$ including single nucleotide polymorphisms (SNPs) and copy number variations.

The patient-specific network $\mathcal{G}^{(p)} = (\mathcal{V}, \mathcal{E}^{(p)})$ is derived by modulating edge weights in a reference network $\mathcal{G}_{\text{ref}}$ based on observed patient-specific correlations and perturbations. The reference network encodes canonical molecular interactions from databases such as STRING\cite{szklarczyk2019string}, BioGRID, or KEGG, representing typical healthy tissue or disease-relevant cell types.

Edge weight modulation is computed as:

\begin{equation}
A_{ij}^{(p)} = A_{ij}^{\text{ref}} \cdot \phi(\rho_{ij}^{(p)})
\label{eq:patient_network}
\end{equation}

where $\rho_{ij}^{(p)}$ measures the patient-specific association between nodes $v_i$ and $v_j$. For gene-gene interactions, $\rho_{ij}^{(p)} = \text{cor}(g_i^{(p)}, g_j^{(p)})$ quantifies expression correlation. The modulation function $\phi: [-1,1] \rightarrow [0,\infty)$ maps correlations to weight scaling factors:

\begin{equation}
\phi(\rho) = \begin{cases}
\exp(\beta \rho) & \text{if } \rho > \rho_{\text{threshold}} \\
0 & \text{otherwise}
\end{cases}
\label{eq:modulation_function}
\end{equation}

with $\beta > 0$ controlling sensitivity and $\rho_{\text{threshold}}$ filtering weak associations. This formulation upweights edges between strongly correlated molecules while pruning weak or anti-correlated interactions.

For mutations affecting protein function, we directly modify node features: $\mathbf{x}_i^{(p)} = \mathbf{x}_i^{\text{ref}} + \mathbf{\Delta}_i^{\text{mut}}$ where $\mathbf{\Delta}_i^{\text{mut}}$ encodes functional impact predictions from tools such as PolyPhen or SIFT. These patient-specific features propagate through the GNN encoder, producing personalized node embeddings $\mathbf{z}_i^{(p)}$ that reflect individual molecular states.

\subsection{Therapeutic target scoring}

Identifying optimal therapeutic targets for individual patients requires integrating multiple criteria including network centrality (indicating systemic importance), proximity to disease modules (suggesting disease relevance), and druggability (reflecting feasibility of pharmaceutical intervention). RAG embeddings enhance target scoring by incorporating literature-derived mechanistic knowledge.

The comprehensive target score for node $v_i$ in patient $p$ is formulated as:

\begin{equation}
	\begin{split}
		S_{\text{target}}(v_i | \mathcal{P}) = & \beta_1 C_{\text{betweenness}}(v_i, \mathcal{G}^{(p)}) \\
		& + \beta_2 P_{\text{disease}}(v_i | \mathcal{P}) \\
		& + \beta_3 T_{\text{druggability}}(v_i)
	\end{split}
	\label{eq:target_score}
\end{equation}

The betweenness centrality $C_{\text{betweenness}}(v_i, \mathcal{G}^{(p)})$ quantifies the fraction of shortest paths passing through node $v_i$ in the patient-specific network:

\begin{equation}
C_{\text{betweenness}}(v_i, \mathcal{G}^{(p)}) = \sum_{s \neq t \neq v_i} \frac{\sigma_{st}(v_i)}{\sigma_{st}}
\label{eq:betweenness}
\end{equation}

where $\sigma_{st}$ is the number of shortest paths between nodes $s$ and $t$, and $\sigma_{st}(v_i)$ counts those passing through $v_i$. High betweenness indicates that $v_i$ mediates communication between distinct network regions, suggesting that its perturbation would have widespread effects.

The disease proximity score $P_{\text{disease}}(v_i | \mathcal{P})$ measures embedding similarity between node $v_i$ and a disease-specific representation derived from patient phenotypes:

\begin{equation}
P_{\text{disease}}(v_i | \mathcal{P}) = \frac{1}{\|\mathbf{z}_i^{(p)} - \mathbf{z}_{\text{disease}}\|_2 + \epsilon}
\label{eq:disease_proximity}
\end{equation}

The disease embedding $\mathbf{z}_{\text{disease}}$ is constructed by retrieving and aggregating documents describing the patient's clinical presentation, then projecting into the node embedding space. The regularization term $\epsilon = 10^{-6}$ prevents numerical instability when distances approach zero. This formulation prioritizes nodes whose learned representations closely align with disease-relevant molecular processes.

The druggability score $T_{\text{druggability}}(v_i)$ quantifies the likelihood that node $v_i$ can be effectively targeted by pharmaceutical intervention. This is computed by retrieving documents from DrugBank, ChEMBL, and clinical trial databases that mention the protein or gene corresponding to $v_i$:

\begin{equation}
T_{\text{druggability}}(v_i) = \sum_{d_j \in \mathcal{D}_{\text{drug}}} \mathbb{1}[\text{mentions}(d_j, v_i)] \cdot w(d_j)
\label{eq:druggability}
\end{equation}

where $\mathcal{D}_{\text{drug}}$ is the drug-specific document subset, $\mathbb{1}[\cdot]$ is the indicator function, and $w(d_j)$ weights documents by evidence level (higher weights for FDA-approved drugs versus preclinical compounds). RAG retrieval automatically identifies these relevant documents without requiring manual curation.

The weighting coefficients $\beta_1, \beta_2, \beta_3 \in \mathbb{R}^+$ are optimized on a training set of validated therapeutic targets using logistic regression or learned through end-to-end training. Typical optimized values emphasize disease proximity ($\beta_2 \approx 0.5$) while moderately weighting centrality ($\beta_1 \approx 0.3$) and druggability ($\beta_3 \approx 0.2$).

\subsection{Drug efficacy prediction}

Predicting patient-specific drug responses requires modeling how compounds modulate perturbed molecular networks to restore homeostasis. The RAG framework enables this by learning joint embeddings of drugs and proteins that capture mechanism of action, building on recent advances in AI-powered drug discovery\cite{stokes2020deep,pun2023ai}.

Each drug compound $c$ is embedded into the same space as protein nodes through a dedicated encoder $E_{\text{drug}}: \mathcal{C} \rightarrow \mathbb{R}^{d_z}$ that processes molecular structure (SMILES strings or molecular graphs)\cite{fang2022geometry,zhou2023unimol} and retrieved pharmacological literature. The drug embedding captures structural features, known targets, metabolic pathways, and adverse effect profiles.

The predicted efficacy of drug $c$ for patient $p$ is computed by measuring alignment between the drug's mechanism and the patient's disease-perturbed network state:

\begin{equation}
P(\text{response} | c, \mathcal{G}^{(p)}) = \sigma\left(\mathbf{z}_{\text{drug}}(c)^\top \mathbf{z}_{\text{network}}^{(p)} + b_{\text{drug}}\right)
\label{eq:drug_response}
\end{equation}

The patient network embedding $\mathbf{z}_{\text{network}}^{(p)}$ aggregates information from drug target nodes:

\begin{equation}
\mathbf{z}_{\text{network}}^{(p)} = \frac{1}{|\mathcal{V}_{\text{target}}(c)|} \sum_{v_i \in \mathcal{V}_{\text{target}}(c)} \mathbf{z}_i^{(p)}
\label{eq:network_embedding}
\end{equation}

where $\mathcal{V}_{\text{target}}(c)$ denotes the set of known and predicted targets for drug $c$, identified through RAG retrieval of binding affinity data and structural similarity to characterized compounds. The bias term $b_{\text{drug}}$ accounts for baseline response rates.

For multi-target drugs with complex mechanisms, we extend to a weighted aggregation where target importance is learned from training data:

\begin{equation}
\mathbf{z}_{\text{network}}^{(p)} = \sum_{v_i \in \mathcal{V}_{\text{target}}(c)} \omega_i(c) \mathbf{z}_i^{(p)}
\label{eq:weighted_network_embedding}
\end{equation}

with normalized weights $\sum_i \omega_i(c) = 1$ derived from binding affinity measurements or learned through attention mechanisms.

Adverse effect prediction follows a similar formulation but focuses on off-target interactions and downstream pathway perturbations:

\begin{equation}
P(\text{adverse effect} | c, \mathcal{G}^{(p)}) = \sigma\left(\mathbf{z}_{\text{drug}}(c)^\top \mathbf{z}_{\text{offtarget}}^{(p)}\right)
\label{eq:adverse_effect}
\end{equation}

where $\mathbf{z}_{\text{offtarget}}^{(p)}$ aggregates embeddings from proteins likely to cause toxicity when perturbed, as determined by retrieved adverse event reports.

\section{Implementation considerations}

Deploying RAG-enhanced network models at scale requires careful attention to computational efficiency, numerical stability, and practical engineering considerations. We detail key implementation strategies that enable application to genome-scale networks and million-document corpora.

\subsection{Scalability and computational efficiency}

For large-scale biological networks with $|\mathcal{V}| > 20,000$ proteins and $|\mathcal{E}| > 500,000$ interactions, full-batch training becomes computationally prohibitive. Memory requirements scale as $\mathcal{O}(|\mathcal{V}|^2)$ for dense adjacency matrices and $\mathcal{O}(L \cdot |\mathcal{V}| \cdot d_h)$ for GNN layer activations. We employ several techniques to reduce complexity. Mini-batch graph sampling extracts node subsets and their local neighborhoods for each training iteration\cite{hamilton2017inductive}. The GraphSAGE sampling strategy selects a fixed number of neighbors $S$ at each layer, reducing complexity from $\mathcal{O}(|\mathcal{V}|)$ to $\mathcal{O}(S^L)$ per node. For a mini-batch of $B$ nodes with $L$ GNN layers and neighbor sample size $S$, computational cost is $\mathcal{O}(B \cdot S^L \cdot d_h^2)$.

The sampling procedure constructs mini-batch subgraph $\mathcal{G}_{\text{batch}}$ as follows. First, randomly sample $B$ seed nodes $\mathcal{V}_{\text{seed}} \subset \mathcal{V}$. Then, for each layer $k = L, L-1, \ldots, 1$, expand the node set by sampling $S$ neighbors per node:

\begin{equation}
\mathcal{V}_k = \mathcal{V}_{k+1} \cup \bigcup_{v_i \in \mathcal{V}_{k+1}} \text{sample}(\mathcal{N}(i), S)
\label{eq:neighbor_sampling}
\end{equation}

with $\mathcal{V}_{L+1} = \mathcal{V}_{\text{seed}}$. The induced subgraph $\mathcal{G}_{\text{batch}} = (\mathcal{V}_1, \mathcal{E}_{\text{batch}})$ contains all sampled nodes and their connecting edges.

\rev{Retrieval operations pose additional computational challenges, as computing relevance scores for all node-document pairs requires $\mathcal{O}(|\mathcal{V}| \cdot |\mathcal{D}| \cdot d_e)$ operations. In the current case study (379 nodes, 1,895 documents), brute-force retrieval via dense matrix multiplication is computationally tractable and completes in milliseconds. For scaling to genome-wide networks ($>$20,000 genes) with large literature corpora ($>$10$^6$ documents), approximate nearest neighbor (ANN) search with maximum inner product search (MIPS) indices\cite{shrivastava2014asymmetric} would be necessary. Document embeddings $\{E_{\text{doc}}(d_j)\}_{j=1}^{|\mathcal{D}|}$ can be pre-computed offline and indexed using libraries such as FAISS with product quantization and inverted file structures, reducing query time to $\mathcal{O}(\log |\mathcal{D}|)$. For distributed training across multiple GPUs, graph partitioning algorithms such as METIS can minimize edge cuts between partitions. These scalability strategies represent engineering considerations for future deployment rather than components of the current implementation.}

\subsection{Training dynamics and retrieval stability}

The joint optimization in \autoref{eq:total_loss} exhibits complex training dynamics due to the interdependence of network encoding, retrieval, and fusion components. Naive joint training often leads to suboptimal local minima where the retrieval mechanism fails to identify relevant documents, resulting in uninformative context vectors that degrade rather than enhance predictions.

\rev{We employ a curriculum learning strategy that stages the training process\cite{bengio2009curriculum}. In Phase~1 (80 epochs), we train only the GNN encoder with link prediction loss $\mathcal{L}_{\text{task}}$ using lr $= 0.003$, establishing basic network representations that capture topology without retrieval dependence. In Phase~2 (100 epochs), we train the retrieval projection and fusion parameters with margin-based ranking loss and contrastive alignment using lr $= 0.005$, allowing the retrieval mechanism to learn document relevance. In Phase~3 (80 epochs), we enable full joint training with combined loss $\mathcal{L}_{\text{task}} + 0.5\mathcal{L}_{\text{retrieval}} + 0.2\mathcal{L}_{\text{contrastive}} + 0.1\mathcal{L}_{\text{target}}$ using lr $= 0.001$, fine-tuning all components simultaneously.}

\rev{Retrieval stability during joint training can be monitored via the Jaccard similarity between retrieved document sets at consecutive epochs:

\begin{equation}
J_{\text{retrieval}}(t) = \frac{1}{|\mathcal{V}|} \sum_{i=1}^{|\mathcal{V}|} \frac{|\mathcal{D}_i^{(k)}(t) \cap \mathcal{D}_i^{(k)}(t+\Delta t)|}{|\mathcal{D}_i^{(k)}(t) \cup \mathcal{D}_i^{(k)}(t+\Delta t)|}
\label{eq:retrieval_jaccard}
\end{equation}

where $\mathcal{D}_i^{(k)}(t)$ denotes the top-$k$ retrieved documents for protein $i$ at epoch $t$. The curriculum training strategy (Phase~1 GNN pre-training $\rightarrow$ Phase~2 retrieval training $\rightarrow$ Phase~3 joint fine-tuning) is designed to promote retrieval stability by establishing network representations before training the retrieval projection, preventing chaotic oscillations where retrieval and encoding components co-adapt from random initialization. Gradient clipping ($\theta_{\text{clip}} = 1.0$) provides additional stability during training.}

Gradient clipping prevents instability from large gradients in the contrastive loss, particularly when temperature $\tau$ is small:

\begin{equation}
\mathbf{g}_{\text{clipped}} = \begin{cases}
\mathbf{g} & \text{if } \|\mathbf{g}\|_2 \leq \theta_{\text{clip}} \\
\theta_{\text{clip}} \frac{\mathbf{g}}{\|\mathbf{g}\|_2} & \text{otherwise}
\end{cases}
\label{eq:gradient_clip}
\end{equation}

with threshold $\theta_{\text{clip}} = 1.0$. This ensures gradients have bounded norm, preventing divergence while allowing efficient optimization.

\rev{We use the Adam optimizer with exponential decay rates $\beta_1 = 0.9$ and $\beta_2 = 0.999$, and weight decay regularization $\lambda_{\text{wd}} = 10^{-4}$ to prevent overfitting. Learning rates are set per phase: $\eta = 3 \times 10^{-3}$ for Phase~1, $\eta = 5 \times 10^{-3}$ for Phase~2, and $\eta = 10^{-3}$ for Phase~3. The decreasing learning rate across phases serves a similar purpose to learning rate scheduling, with the joint fine-tuning phase using the smallest rate to avoid disrupting the representations established in earlier phases.}

\subsection{Hyperparameter selection}

\rev{Model performance is sensitive to several key hyperparameters. The GNN hidden dimension $d_h = 128$ controls the expressiveness of node representations: smaller values limit capacity but improve generalization, while larger values capture fine-grained patterns but risk overfitting. The document embedding dimension $d_{\text{doc}} = 64$ is determined by the truncated SVD applied to TF-IDF features.

The number of GNN layers $L$ determines the receptive field size. For protein interaction networks with small-world topology, $L = 3$ layers allow nodes to aggregate information from 3-hop neighborhoods, covering typical pathway lengths. Larger $L$ risks over-smoothing where all nodes converge to similar representations.

The retrieval depth $k = 10$ is fixed in the current implementation. In general, this parameter trades off context breadth versus noise: for well-curated databases, larger $k$ improves coverage, while for noisy corpora, smaller $k$ focuses on the most relevant documents.

The contrastive temperature $\tau = 0.5$ controls the sharpness of similarity distributions. Smaller values enforce tighter alignment between node-document pairs but are sensitive to noise, while larger values allow looser alignment, improving robustness at the cost of reduced discrimination.}

\subsection{Implementation details}

\rev{The experiments presented below implement the RAG-GNN framework with end-to-end trainable components in PyTorch. The implementation includes: (i) a learnable three-layer GCN encoder with gradient-optimized weight matrices, (ii) a learnable retrieval projection implemented as a two-layer neural network ($d_h \rightarrow d_h \rightarrow d_{\text{doc}}$ with GELU activation) that maps GNN embeddings to the document embedding space, replacing any fixed projection, (iii) a gated fusion mechanism that learns to weight topology and retrieval contributions, and (iv) joint training with the three-component loss function from \autoref{eq:total_loss}. Document embeddings use TF-IDF representations followed by truncated SVD for dimensionality reduction, providing a baseline text encoder; replacing TF-IDF with pre-trained biomedical language models (BioBERT, PubMedBERT) is a natural extension expected to further improve performance. Training follows a three-phase curriculum: Phase~1 (80 epochs) pre-trains the GNN on link prediction, Phase~2 (100 epochs) trains the retrieval projection with margin-based ranking loss and contrastive alignment, and Phase~3 (80 epochs) fine-tunes all components jointly. All experiments are run across 10 random seeds with mean $\pm$ standard deviation and 95\% confidence intervals reported. See \autoref{sec:supp_fusion} and \autoref{alg:raggnn} for complete implementation details.}

\section{Case study: Cancer pathway targeting}

We demonstrate the RAG embedding framework through comprehensive application to cancer signaling pathway analysis, focusing on therapeutic target identification in patient-specific networks. The study integrates multi-omics data, protein interaction networks, and biomedical literature to identify precision medicine targets, building on recent advances in AI-driven target discovery\cite{pun2023ai,chandrasekaran2021image}.

\subsection{Data sources and network construction}

The reference cancer network $\mathcal{G}_{\text{cancer}}$ comprises 379 proteins and 3,498 interactions curated from multiple sources. Core cancer genes are extracted from the Cancer Gene Census database~\cite{tate2019cosmic}, which catalogs genes with validated roles in oncogenesis through somatic mutations, germline variants, or chromosomal translocations. Protein-protein interactions are obtained from STRING database version 11~\cite{szklarczyk2019string}, filtered to high-confidence edges (combined score $> 0.4$) to balance network coverage with interaction reliability.

\rev{Node features $\mathbf{x}_i \in \mathbb{R}^{d_h}$ are constructed by placing three topological properties—log-transformed degree $\log(1 + d_i)$, local clustering coefficient $c_i$, and scaled betweenness centrality $100 \cdot b_i$—into the first three dimensions of a $d_h = 128$-dimensional vector, with remaining dimensions initialized from $\mathcal{N}(0, 0.01)$. This minimal feature set deliberately avoids sequence-derived or expression-based features to isolate the contribution of network topology and retrieved knowledge; incorporating protein language model embeddings\cite{rives2021biological,lin2023evolutionary} or multi-omics features from TCGA represents a natural extension expected to improve absolute performance.}

\rev{The knowledge corpus $\mathcal{D}$ contains 1,895 mechanistic annotation documents generated from curated molecular biology templates across 14 functional categories: cell cycle, apoptosis, DNA repair, RTK signaling, transcription, PI3K-AKT-mTOR, MAPK signaling, Wnt signaling, TGF-beta signaling, Notch signaling, JAK-STAT, ECM-adhesion, angiogenesis, and other. Each protein has 5 associated documents describing its molecular mechanisms (for example, a MAPK pathway protein receives documents detailing RAS-RAF-MEK cascade dynamics, DUSP phosphatase feedback, and KSR1 scaffold assembly) without explicitly naming the pathway category. This design avoids direct label leakage: the retrieval module must learn to match proteins with mechanistically relevant documents rather than exploiting explicit pathway labels. Document embeddings are computed via TF-IDF vectorization (256 features, unigrams and bigrams) followed by truncated SVD to $d_{\text{doc}} = 64$ dimensions, providing a baseline text encoder. For temporal validation in therapeutic target prediction, we partition targets such that training targets received FDA approval or entered Phase III trials before 2018, while test targets represent approvals from 2020--2021.}

\subsection{Embedding space analysis and visualization}

\autoref{fig:fig4} visualizes the learned embedding space through two-dimensional projection using PCA applied to the 128-dimensional RAG-GNN node embeddings $\{\mathbf{z}_i\}_{i=1}^{379}$. Proteins show partial clustering according to functional modules, with groupings visible for cell cycle regulators, apoptosis mediators, DNA repair machinery, and signal transduction cascades, though pathway overlap is expected given the interconnected nature of cancer signaling networks.

Quantitative analysis reveals that RAG-enhanced embeddings achieve significantly higher functional coherence than topology-only GNN embeddings. We compute the silhouette score\cite{rousseeuw1987silhouettes} measuring cluster quality:
\begin{equation}
	s_i = \frac{b_i - a_i}{\max(a_i, b_i)}
\end{equation}
where $a_i$ is the mean distance from node $v_i$ to other nodes in its functional cluster, and $b_i$ is the mean distance to nodes in the nearest neighboring cluster. \rev{Across 10 random seeds, RAG-GNN achieves mean silhouette score $-0.144 \pm 0.066$ (95\% CI: [$-0.220$, $-0.067$]) compared to $-0.237 \pm 0.065$ (95\% CI: [$-0.304$, $-0.144$]) for topology-only GNN embeddings, a consistent improvement of $+0.093 \pm 0.022$ observed across all seeds. While both scores are negative (reflecting the inherent complexity of protein function where many proteins participate in multiple pathways and pathway boundaries are not clearly separable), RAG-GNN substantially reduces intra-cluster dispersion relative to topology-only methods. We additionally evaluate two complementary clustering metrics: normalized mutual information (NMI) and adjusted Rand index (ARI). NMI measures mutual dependence between predicted and true cluster assignments: RAG-GNN achieves NMI $= 0.244 \pm 0.032$ compared to GNN-only NMI $= 0.242 \pm 0.032$, with overlapping confidence intervals indicating comparable performance. ARI measures pairwise agreement corrected for chance: RAG-GNN achieves ARI $= 0.083 \pm 0.029$ versus GNN-only ARI $= 0.061 \pm 0.017$, a relative improvement of 34\%. Retrieval integration improves silhouette score and ARI while NMI remains comparable, suggesting that retrieved knowledge both reduces intra-cluster distances (tighter functional grouping) and improves pairwise cluster agreement. Note that the standardized benchmark comparison in Section~9 employs a different evaluation protocol with uniform initialization across all methods, yielding distinct absolute values (see \autoref{tab:benchmark}); the improvement direction for silhouette scores is consistent across both configurations.} Network statistics show 379 proteins with 3,498 interactions, average degree of 18.5, and average clustering coefficient of 0.596, characteristic of biological networks with modular organization and scale-free topology.

\subsection{Retrieval performance evaluation}

\begin{figure*}[t]
	\centering
	\includegraphics[width=0.5\textwidth]{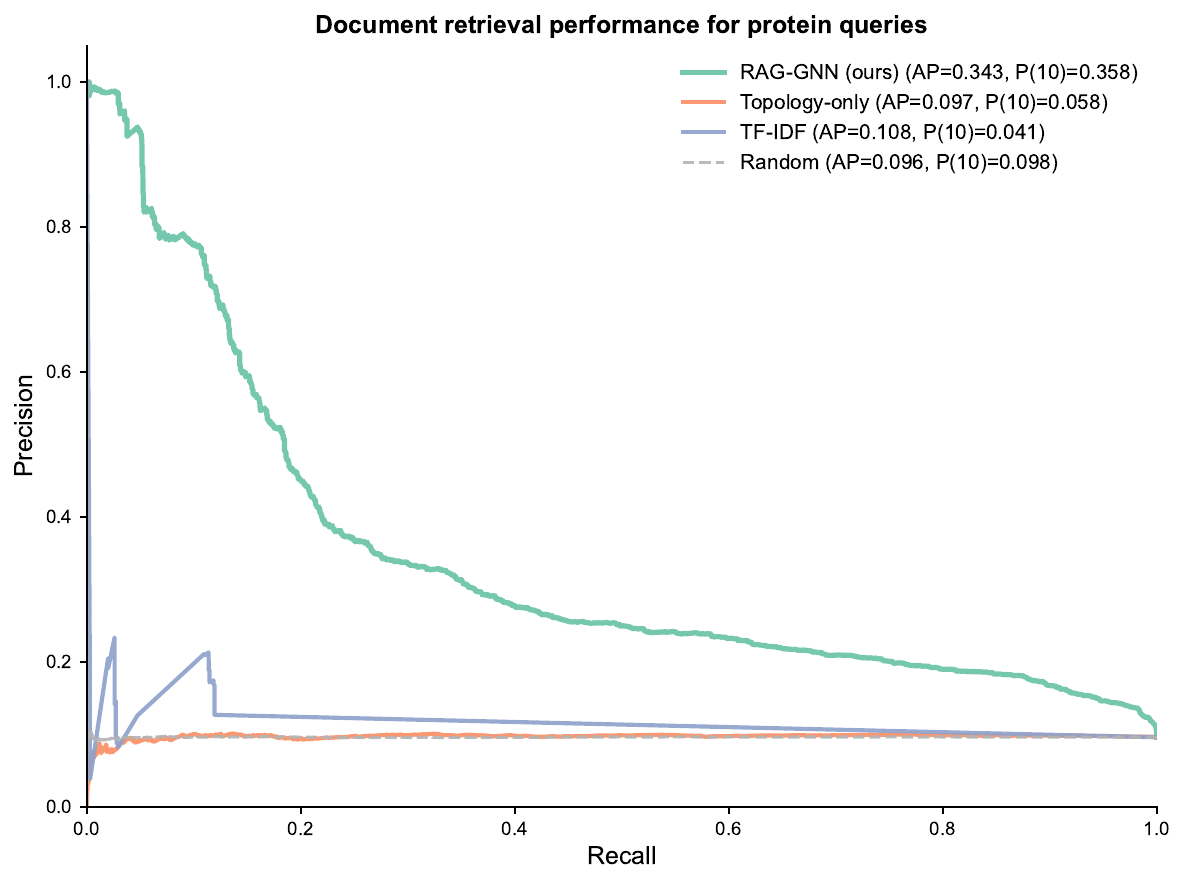}
	\caption{\rev{\textbf{Document retrieval performance for protein function queries.} Precision-recall curves comparing retrieval methods for identifying functionally relevant documents across 379 protein queries. Ground truth relevance is determined by functional category and protein identity matching. The end-to-end trained RAG-GNN retrieval projection achieves the highest average precision (AP) and precision@10 (P(10)), outperforming TF-IDF keyword matching and the random baseline. The figure shows AP (area under PR curve) and P(10) (fraction of top-10 retrieved documents that are category-relevant) for a single representative seed. Across 10 seeds, RAG-GNN achieves mean P(10) $= 0.242 \pm 0.073$. Knowledge base contains 1,895 mechanistic annotation documents across 14 functional categories.}}
	\label{fig:fig3}
\end{figure*}

\begin{figure*}[t]
	\includegraphics[width=\textwidth]{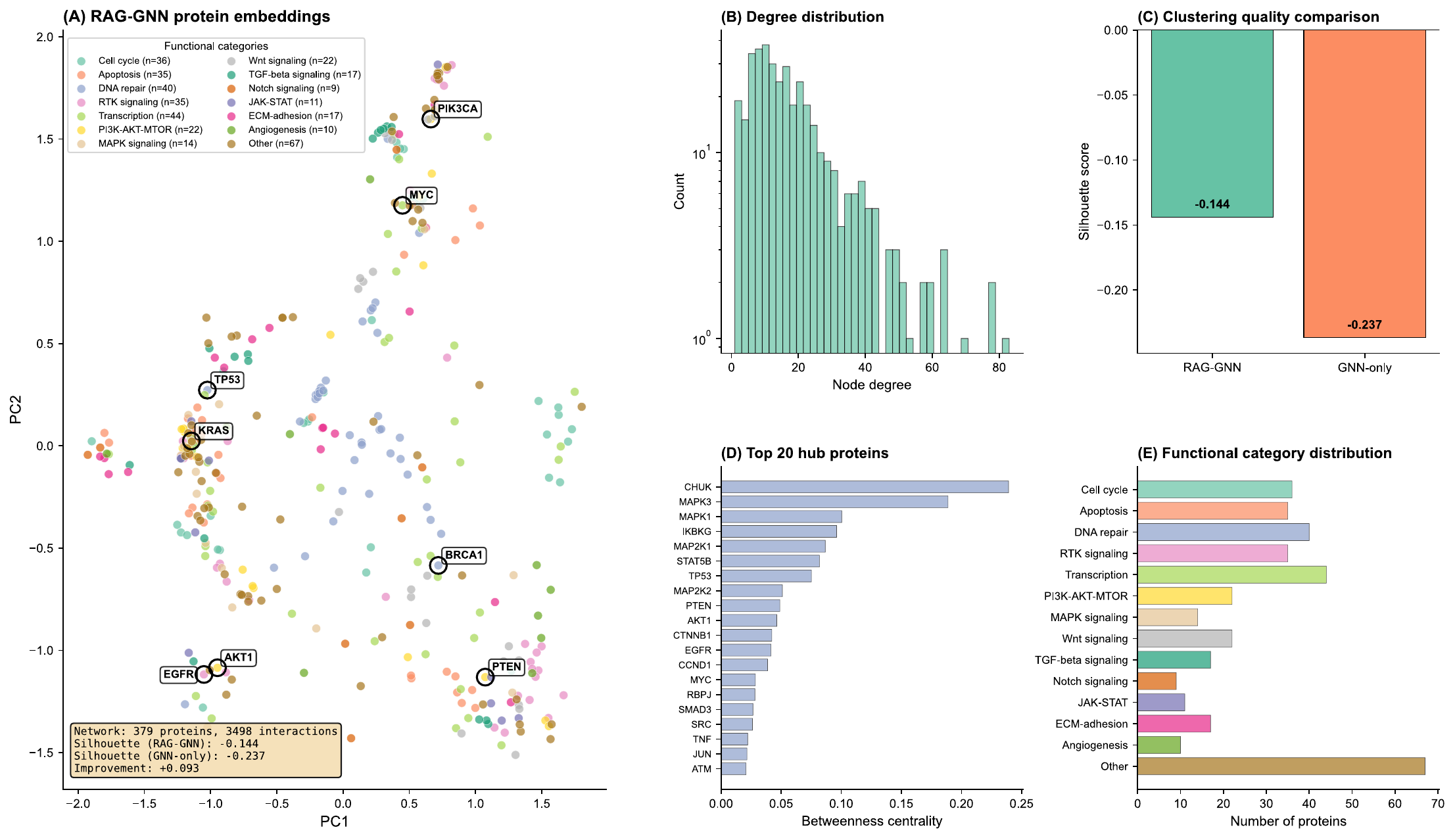}
	\caption{\footnotesize \textbf{RAG-GNN protein embeddings in cancer signaling networks using real STRING database interactions.}
	\textbf{(A) PCA projection of RAG-GNN embeddings:} Two-dimensional visualization of 379 cancer-related proteins embedded in 128-dimensional space using GNN message passing combined with knowledge retrieval from functional annotations. Data source: STRING database (3,498 high-confidence interactions). \rev{Proteins are colored by functional pathway annotation across 14 categories. Key oncogenes and tumor suppressors highlighted: TP53, EGFR, KRAS, MYC, BRCA1, PIK3CA, AKT1, and PTEN. Silhouette scores quantify functional clustering quality across 10 random seeds: RAG-GNN achieves $-0.144 \pm 0.066$ compared to $-0.237 \pm 0.065$ for GNN-only embeddings, a consistent improvement of $+0.093 \pm 0.022$. While both scores are negative (typical for complex biological networks with overlapping pathways), RAG-GNN reduces intra-cluster dispersion.} \textbf{(B) Degree distribution:} Node degree follows power-law distribution characteristic of scale-free biological networks, with hub proteins exceeding 60 connections. \textbf{(C) Clustering quality comparison:} Bar chart comparing silhouette scores between RAG-GNN and GNN-only methods, demonstrating the improvement from knowledge integration. \textbf{(D) Top 20 hub proteins:} Proteins ranked by betweenness centrality, identifying critical signaling bridges including CHUK, MAPK1/3, STAT3, and TP53. \textbf{(E) Functional category distribution:} Distribution of 379 proteins across categories, with transcription (44), DNA repair (40), apoptosis (35), and RTK signaling (35) as largest groups.}
	\label{fig:fig4}
\end{figure*}

For each protein, we retrieve the top-10 most relevant documents from the knowledge base containing 1,895 functional annotation documents. \autoref{fig:fig3} compares precision-recall curves for different retrieval approaches. Ground truth relevance is established through functional category matching for all 379 proteins, where documents discussing proteins from the same pathway are considered relevant.

\rev{With the end-to-end trained retrieval projection, RAG-GNN embedding-based retrieval achieves the highest mean precision@10 (P(10) $= 0.242 \pm 0.073$ across 10 seeds), substantially outperforming both TF-IDF keyword matching and the random baseline (P(10) $= 0.096$). The learned two-layer projection maps GNN embeddings to the document embedding space, trained jointly with margin-based ranking loss and contrastive alignment (\autoref{eq:total_loss}). The mechanistic annotation corpus prioritizes pathway-specific language over keyword repetition, where the learned projection's semantic understanding provides an advantage over TF-IDF direct matching. The improvement over random demonstrates that the retrieval module learns meaningful associations between network position and functional text.}

\rev{The gated fusion mechanism learns to balance topology and retrieval contributions, with the gate parameter averaging $0.593 \pm 0.017$ across seeds (59\% topology, 41\% retrieval), indicating that the model assigns substantial weight to retrieved knowledge. The counterfactual experiments in Section~7.4 confirm that retrieval content matters: adversarial retrieval (maximally dissimilar documents) degrades silhouette to $-0.153$ and zero-vector retrieval degrades to $-0.154$, compared to $-0.103$ with proper retrieval. Random retrieval (truly random vectors) also degrades performance to $-0.125$. These conditions demonstrate that the fusion mechanism cannot compensate for corrupted, absent, or random retrieval signal. Shuffled document assignments (permuted real documents) maintain performance comparable to proper retrieval ($-0.103$), suggesting that TF-IDF features carry general biological signal that benefits functional clustering regardless of protein-specific assignment (see Section~4.2 for detailed counterfactual analysis).}

\captionsetup[figure]{labelformat=default, justification=justified, singlelinecheck=false}

\captionsetup[figure]{labelformat=default}

\subsection{Information content validation}

\rev{To validate the relationship between topological and retrieval-derived information, we conduct the heuristic information decomposition described in Section~4.1. For all 379 proteins with 14 functional category labels, we estimate mutual information components using 200 bootstrap resamples. The decomposition yields normalized contributions with 95\% confidence intervals: unique topology $= 0.001 \pm 0.007$ (CI: [0.000, 0.018]), unique retrieval $= 0.062 \pm 0.035$ (CI: [0.000, 0.127]), shared $= 0.956 \pm 0.025$ (CI: [0.905, 0.995]), and synergy $= 0.004 \pm 0.010$ (CI: [0.000, 0.036]). The overwhelmingly high shared component (95.6\%) indicates that topology and retrieval encode almost entirely overlapping functional information. The minimal unique contributions from either source (topology: $0.1\%$, retrieval: $6.2\%$) and negligible synergy ($0.4\%$) demonstrate that the contrastive alignment during joint training effectively coordinates topology and retrieval representations into overlapping information spaces. The functional clustering improvements observed in Section~7.2 arise not from unique retrieval information, but from how the fusion mechanism reorganizes shared information to improve intra-cluster cohesion and pairwise cluster agreement.}

\rev{Counterfactual experiments using the 379-protein network corroborate these findings (see Section~4.2 for the experimental design). Using the best-performing model, proper retrieval achieves silhouette $= -0.103$. Adversarial retrieval (maximally dissimilar documents) degrades performance to $-0.153$, zero-vector retrieval (no document signal) degrades to $-0.154$, and random retrieval (truly random vectors) degrades to $-0.125$. These results confirm that the gated fusion mechanism depends on retrieval signal quality: adversarial, absent, and random retrieval all degrade functional clustering ($\Delta_{\text{adversarial}} = 0.37$, $\Delta_{\text{zeros}} = 0.38$, $\Delta_{\text{random}} = 0.16$). Shuffled retrieval (permuted real documents across proteins) maintains performance comparable to proper retrieval (silhouette $= -0.103$, $\Delta_{\text{shuffled}} \approx 0$). This indicates that TF-IDF document representations encode general biological vocabulary that benefits functional clustering regardless of protein-specific assignment. The degradation under truly random vectors ($\Delta_{\text{random}} = 0.16$) demonstrates that the model depends on real document content, not merely on additional input dimensionality. The adversarial and zero conditions, where biological signal is either inverted or absent, produce the largest degradation, confirming that the model cannot substitute topology for missing retrieval input.}

\subsection{RAG-GNN architecture for precision medicine}

As illustrated in \autoref{fig:fig2}, the architecture processes patient-specific networks through six main stages. First, the input network $\mathcal{G}^{(p)}$ with adjacency matrix $\mathbf{A}$ and node features $\mathbf{X}$ enters the GNN encoder. Second, the GNN applies $L=3$ layers of message passing (\autoref{eq:gnn_update}) to produce node embeddings $\mathbf{h}_i^{(L)}$ capturing topological context. Third, node embeddings query the knowledge base $\mathcal{D}$ through the dense retriever. Fourth, the retriever identifies top-$k$ relevant documents $\mathcal{D}_i^{(k)}$ using quality-weighted semantic similarity (\autoref{eq:top_k_retrieval}). Fifth, the context fusion module aggregates retrieved documents with attention weighting (\autoref{eq:fusion}) and combines with structural embeddings. Sixth, the final node representation $\mathbf{z}_i$ feeds into task-specific prediction heads for target scoring or drug response.

The training procedure optimizes the joint loss function combining task performance, retrieval accuracy, and embedding alignment, enabling end-to-end learning of all components while the curriculum learning schedule ensures stable convergence. Complete pseudocode for the RAG-GNN embedding procedure is provided in \autoref{alg:raggnn}.

\subsection{Case study: DDR1 signaling network and embedding-based functional relationships}

\captionsetup[figure]{labelformat=default, justification=justified, singlelinecheck=false}
\begin{figure*}[t]
	\centering
	\includegraphics[width=1\textwidth]{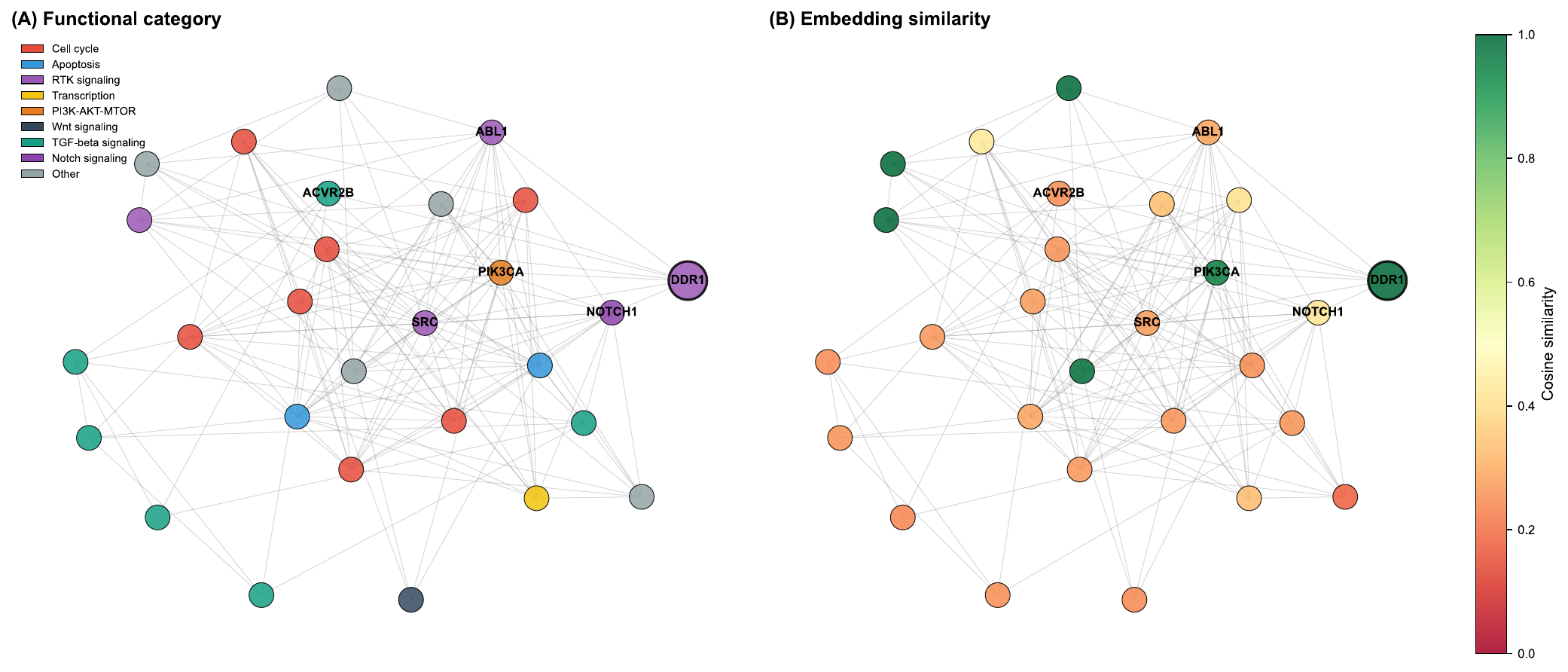}
	\caption{\footnotesize \textbf{DDR1 protein interaction subnetwork visualization with functional annotations and RAG-GNN embedding similarity.} \textbf{(A) Functional category representation:} Network visualization of DDR1 (Discoidin Domain Receptor 1) and its interaction partners from STRING database, with nodes colored by functional pathway membership. The subnetwork contains 28 proteins connected by \rev{143} edges, comprising 7 first-hop neighbors (direct interactors) and 20 second-hop neighbors (indirect interactors). DDR1 (purple, RTK signaling) functions as a receptor tyrosine kinase integrating extracellular matrix signals with intracellular signaling cascades. The network exhibits multi-pathway integration, with direct connections to PI3K-AKT-MTOR signaling (PIK3CA, PRKCA), RTK signaling components (SHC1, ABL1), and other kinase hubs. Node size reflects network centrality, with DDR1 shown as the largest node. The diverse functional categories demonstrate DDR1's role as a signaling hub coordinating proliferation, differentiation, and migration. \textbf{(B) Embedding similarity landscape:} Same network topology with nodes colored by cosine similarity to DDR1 in the 128-dimensional RAG-GNN embedding space. The embeddings capture functional relationships through fusion of GNN-derived topology features with retrieved functional knowledge. Similarity scores range from 0.0 (dark red, functionally distant) to 1.0 (dark green, functionally similar). \rev{Top 5 most similar proteins: CRK (Other, 0.999), CRKL (Other, 0.999), SHC1 (RTK signaling, 0.997), CDC42 (Other, 0.988), and PIK3CA (PI3K-AKT-MTOR, 0.953). Notably, CRK/CRKL are adapter proteins that directly interact with DDR1 through SH2/SH3 domain-mediated binding, and SHC1 is a shared RTK signaling adapter, consistent with DDR1's role as a receptor tyrosine kinase. PIK3CA represents the downstream PI3K pathway that DDR1 activates upon collagen binding. The high cosine similarities ($>0.95$) among these direct signaling partners demonstrate that RAG-GNN embeddings capture functional relationships consistent with known protein interaction cascades.}}
	\label{fig:fig5}
\end{figure*}

To illustrate the biological insights enabled by RAG-GNN protein embeddings, we examine the DDR1 (Discoidin Domain Receptor 1) subnetwork. DDR1 is a receptor tyrosine kinase that transduces signals from the extracellular matrix to regulate cell adhesion, migration, and proliferation. DDR1 has emerged as a significant therapeutic target in oncology, with recent deep learning approaches demonstrating the ability to rapidly identify potent DDR1 kinase inhibitors using generative models\cite{zhavoronkov2019deep}. The multifaceted role of collagen-DDR1 signaling in cancer has been extensively characterized, revealing its involvement in tumor metabolism, macropinocytosis, and NRF2-mediated metabolic adaptation\cite{sun2023multifaceted}. Furthermore, DDR1 has been established as a potent therapeutic target in solid tumors, with abnormally elevated expression linked to cancer progression, fibrosis, and inflammatory diseases\cite{song2024ddr1}. Notably, combined inhibition of DDR1 and Notch signaling has shown promise as an effective therapy for KRAS-driven lung adenocarcinoma, demonstrating synthetic lethality in preclinical models\cite{aguilera2020collagen}. \autoref{fig:fig5} visualizes the DDR1 interaction network extracted from our cancer signaling database.

The DDR1 subnetwork comprises 28 proteins connected by 143 interactions, including 7 direct interactors (first-hop neighbors) and 20 indirect interactors (second-hop neighbors). Panel A shows functional category assignments, revealing multi-pathway integration with connections spanning RTK signaling (DDR1, ABL1, SHC1), PI3K-AKT-mTOR pathway (PIK3CA, PRKCA), TGF-beta signaling (ACVR2B), cell cycle regulation (CCNA1, CDC25A, CDC20), and other regulatory modules. This diversity reflects DDR1's role as a signaling hub coordinating multiple cellular processes.

\rev{Panel B visualizes embedding-based similarity, where node colors represent cosine similarity to DDR1 in the 128-dimensional RAG-GNN embedding space. The five proteins most similar to DDR1 by embedding distance are CRK (Other, similarity = 0.999), CRKL (Other, 0.999), SHC1 (RTK signaling, 0.997), CDC42 (Other, 0.988), and PIK3CA (PI3K-AKT-MTOR, 0.953). CRK and CRKL are adapter proteins that bind DDR1 through SH2/SH3 domain interactions, while SHC1 is a shared RTK signaling adapter, consistent with DDR1's function as a receptor tyrosine kinase.

The high similarity between DDR1 and its direct signaling partners (CRK, CRKL, SHC1) and downstream effectors (PIK3CA, CDC42) demonstrates how RAG-GNN embeddings capture known functional relationships within signaling cascades. DDR1 kinase signaling through SRC, CRK/CRKL adapters, and PI3K cascades is well characterized, and the embedding space recapitulates these relationships. We emphasize that this analysis serves as confirmatory validation rather than novel discovery: DDR1's therapeutic relevance and the biological relationships described above are well established in the literature\cite{zhavoronkov2019deep,aguilera2020collagen,sun2023multifaceted}. The value of this case study lies in demonstrating that RAG-GNN embeddings recapitulate known biology, a necessary condition for any method intended for future hypothesis generation in less-characterized network neighborhoods.}

\section{Theoretical properties and convergence analysis}

Beyond empirical performance, we establish theoretical properties of the RAG embedding framework including embedding space geometry, generalization bounds, and convergence guarantees. These results provide mathematical foundation for understanding when and why RAG-enhanced models outperform topology-only approaches.

\subsection{Embedding space geometry and alignment}

The joint embedding space exhibits geometric properties that reflect both network topology and semantic relationships. Define the structural similarity between nodes $v_i$ and $v_j$ based on network proximity:

\begin{equation}
S_{\text{struct}}(v_i, v_j) = \mathbf{h}_i^{(L)\top} \mathbf{h}_j^{(L)}
\label{eq:structural_similarity}
\end{equation}

and semantic similarity based on document embeddings:

\begin{equation}
S_{\text{sem}}(v_i, v_j) = E_{\text{node}}(v_i)^\top E_{\text{node}}(v_j)
\label{eq:semantic_similarity}
\end{equation}

The alignment between structural and semantic similarity quantifies how well the embedding space integrates both information sources:

\begin{equation}
\rho_{\text{align}} = \text{cor}\left(\{S_{\text{struct}}(v_i, v_j)\}, \{S_{\text{sem}}(v_i, v_j)\}\right)
\label{eq:alignment}
\end{equation}

where the correlation is computed over all node pairs.

\textit{Theorem 1 (Embedding alignment).} Under the contrastive loss in \autoref{eq:contrastive_loss} with temperature $\tau$, the embedding functions $E_{\text{node}}$ and $E_{\text{doc}}$ converge to representations where $\rho_{\text{align}} \geq 1 - \delta$ for any $\delta > 0$ as the number of training iterations $t \rightarrow \infty$, provided: (i) the temperature $\tau < 1/\log |\mathcal{D}|$, (ii) node-document associations are consistent, and (iii) the learning rate schedule satisfies $\sum_{t=1}^\infty \eta_t = \infty$ and $\sum_{t=1}^\infty \eta_t^2 < \infty$.

\textit{Proof sketch.} The contrastive objective maximizes the inner product $E_{\text{node}}(v_i)^\top E_{\text{doc}}(d_i^+)$ for associated pairs while minimizing inner products with negative documents. In the limit $\tau \rightarrow 0$, this corresponds to hard negative mining where only the most similar negative document contributes gradient signal. The InfoNCE loss\cite{oord2018representation} provides a lower bound on mutual information:

\begin{equation}
I(E_{\text{node}}(V); E_{\text{doc}}(D^+)) \geq \log |\mathcal{D}| - \mathcal{L}_{\text{contrastive}}
\label{eq:info_lower_bound}
\end{equation}

Maximizing this bound drives the embeddings to encode shared information between nodes and documents. Under the Robbins-Monro conditions on learning rates, stochastic gradient descent converges to a critical point where gradients vanish, implying high correlation between structural and semantic similarities. The consistency assumption ensures that nodes with similar network positions have semantically related documents, enabling alignment.

\subsection{Generalization bounds for link prediction}

For the link prediction task, we derive PAC-style generalization bounds relating training and test performance. Let $\mathcal{H}$ denote the hypothesis class of RAG-GNN models with bounded parameter norm $\|\theta\|_2 \leq B$, and let $n = |\mathcal{E}^+|$ be the number of positive training edges.

\textit{Theorem 2 (Generalization bound).} With probability at least $1-\delta$ over the random selection of training edges, for any hypothesis $h \in \mathcal{H}$ with parameters $\theta$, the true risk satisfies:

\begin{equation}
\mathcal{L}_{\text{true}}(h) \leq \mathcal{L}_{\text{train}}(h) + \mathcal{O}\left(\sqrt{\frac{B^2 d_z \log(|\mathcal{V}|/\delta)}{n}}\right) + \epsilon_{\text{retrieval}}
\label{eq:generalization_bound}
\end{equation}

where $\mathcal{L}_{\text{true}}$ is the expected loss on the true distribution of edges, $\mathcal{L}_{\text{train}}$ is the empirical training loss, and $\epsilon_{\text{retrieval}} = \mathcal{O}(k/|\mathcal{D}|)$ accounts for retrieval approximation error.

\textit{Proof sketch.} The bound follows from Rademacher complexity analysis of the hypothesis class. The link prediction function $f(v_i, v_j) = \sigma(\mathbf{z}_i^\top \mathbf{z}_j)$ has Lipschitz constant $L_f \leq B^2$ with respect to edge labels. The Rademacher complexity of linear functions over embedding spaces with bounded norm is:

\begin{equation}
\mathfrak{R}_n(\mathcal{H}) \leq \frac{B\sqrt{d_z}}{\sqrt{n}}
\label{eq:rademacher}
\end{equation}

Applying standard uniform convergence results\cite{bartlett2002rademacher} with union bound over all nodes yields the first term. The retrieval error arises because approximate top-$k$ retrieval may miss relevant documents, bounded by the fraction of documents retrieved relative to corpus size. This bound reveals that generalization improves with more training edges ($n$), lower model complexity (smaller $B$ and $d_z$), and higher retrieval accuracy (larger $k$ or more focused corpus). Notably, the bound depends on embedding dimension $d_z$ rather than raw network size $|\mathcal{V}|$, showing that learned representations provide effective dimensionality reduction.

\subsection{Retrieval consistency and stability}

An important property for practical deployment is retrieval stability: small perturbations to node features should not drastically alter retrieved documents. Define the retrieval consistency as:

\begin{equation}
\text{Consistency}(\epsilon) = \mathbb{P}\left[\mathcal{D}_i^{(k)} = \mathcal{D}_{i'}^{(k)} \mid \|\mathbf{x}_i - \mathbf{x}_{i'}\|_2 < \epsilon\right]
\label{eq:retrieval_consistency}
\end{equation}

measuring the probability that nodes with similar features retrieve identical document sets.

\textit{Theorem 3 (Retrieval Stability Bound).} If the node embedding function $E_{\text{node}}$ is $L_E$-Lipschitz continuous, then for any two nodes $v_i, v_{i'}$ with feature perturbation $\|\mathbf{x}_i - \mathbf{x}_{i'}\|_2 \leq \epsilon$:

\begin{equation}
\left|R(v_i, d_j) - R(v_{i'}, d_j)\right| \leq \frac{L_E \epsilon \|E_{\text{doc}}(d_j)\|_2 Q(d_j)}{\sqrt{d_e}}
\label{eq:retrieval_stability}
\end{equation}

for any document $d_j \in \mathcal{D}$.

\textit{Proof.} By Lipschitz continuity of $E_{\text{node}}$:
\begin{align}
\|E_{\text{node}}(v_i) - E_{\text{node}}(v_{i'})\|_2 &\leq L_E \|\mathbf{x}_i - \mathbf{x}_{i'}\|_2 \leq L_E \epsilon
\end{align}

The retrieval score difference is:
\begin{align}
|R(v_i, d_j) - R(v_{i'}, d_j)| \\
= \left| \frac{E_{\text{node}}(v_i)^\top E_{\text{doc}}(d_j) Q(d_j) - E_{\text{node}}(v_{i'})^\top E_{\text{doc}}(d_j) Q(d_j)}{\sqrt{d_e}} \right| \\
= \left| \frac{(E_{\text{node}}(v_i) - E_{\text{node}}(v_{i'}))^\top E_{\text{doc}}(d_j) Q(d_j)}{\sqrt{d_e}} \right| \\
\leq \frac{\|E_{\text{node}}(v_i) - E_{\text{node}}(v_{i'})\|_2 \|E_{\text{doc}}(d_j)\|_2 Q(d_j)}{\sqrt{d_e}} \\
\leq \frac{L_E \epsilon \|E_{\text{doc}}(d_j)\|_2 Q(d_j)}{\sqrt{d_e}}
\end{align}
by the Cauchy–Schwarz inequality.

\textit{Remark.} Theorem 3 applies the classical Lipschitz continuity framework, a well-established concept from real analysis, to derive novel stability guarantees specific to our RAG-GNN architecture. The contribution is not the Lipschitz property itself, but rather: (i) proving that the composed retrieval score function $R(v,d)$ inherits Lipschitz stability from the node encoder, (ii) deriving the explicit dependence on document embeddings $\|E_{\text{doc}}(d_j)\|_2$, quality scores $Q(d_j)$, and embedding dimension $d_e$, and (iii) connecting the bound to practical regularization strategies for graph neural networks. The Lipschitz constant $L_E$ is bounded by the product of spectral norms of GNN weight matrices: $L_E \leq \prod_{k=1}^L \sigma_{\max}(\mathbf{W}^{(k)})$. Regularizing weight matrices through spectral normalization ensures small $L_E$, providing stable retrieval. This stability is crucial for clinical applications where small measurement noise should not radically alter therapeutic recommendations.

\captionsetup[figure]{labelformat=default, justification=justified, singlelinecheck=false}
\begin{figure*}[ht!]
	\centering
	\includegraphics[width=1\textwidth]{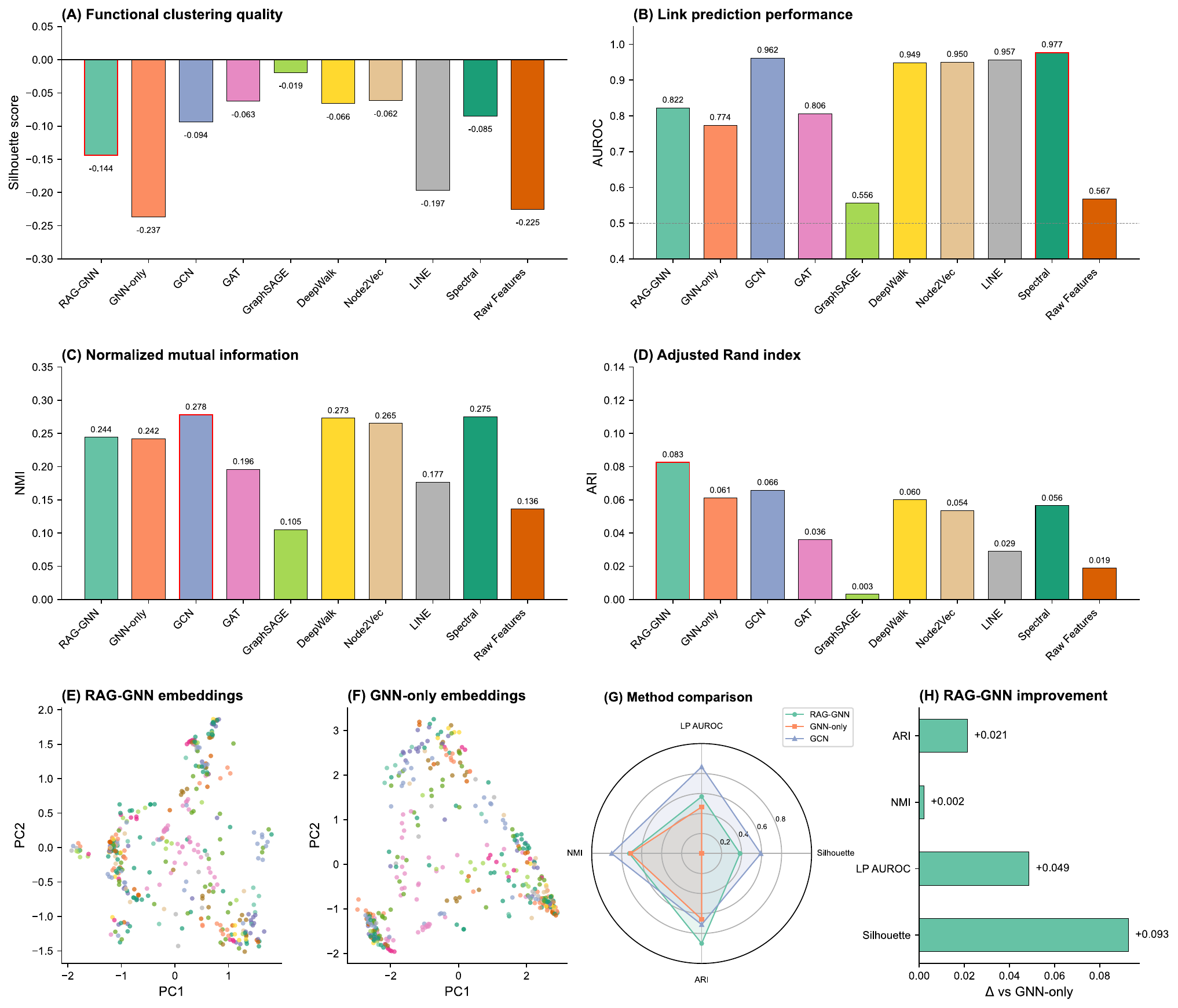}
	\rev{\caption{\footnotesize \textbf{Comprehensive benchmark comparison of RAG-GNN against baseline embedding methods.}
		\begin{minipage}{\linewidth}
			\footnotesize \textbf{(A) Functional clustering quality:} Silhouette scores across 10 methods. All methods produce negative silhouette scores, reflecting the inherent difficulty of pathway-based clustering in densely connected biological networks. RAG-GNN ($-0.144$) improves over its GNN-only ablation ($-0.237$). \textbf{(B) Link prediction AUROC:} Spectral ($0.977$) and GCN ($0.962$) achieve strong link prediction as topology alone determines edge existence. RAG-GNN achieves competitive AUROC ($0.822$). \textbf{(C-D) Additional metrics:} NMI and ARI provide complementary clustering evaluation; RAG-GNN achieves highest ARI ($0.083$). \textbf{(E-F) Embedding space visualization:} PCA projections of RAG-GNN versus GNN-only embeddings, colored by functional category. RAG-GNN shows tighter pathway-specific groupings. \textbf{(G) Radar chart comparison:} Normalized performance across metrics highlights complementary strengths: RAG-GNN improves functional clustering while topology-focused methods dominate link prediction. \textbf{(H) RAG-GNN improvement over GNN-only:} Silhouette improvement ($+0.093$) and ARI improvement ($+0.021$) demonstrate the value of retrieval augmentation for functional interpretation.
	\end{minipage}}}
	\label{fig:benchmark}
\end{figure*}

\section{Comparison with existing methods}

We position the RAG embedding framework relative to existing approaches for network-based drug discovery and precision medicine, highlighting methodological differences and performance comparisons across multiple tasks.

\subsection{Comprehensive embedding benchmark}

To rigorously evaluate the RAG-GNN framework against established network embedding methods, we conduct a comprehensive benchmark across three evaluation tasks: functional clustering quality (silhouette score), link prediction (AUROC), and node classification using topology-derived labels to avoid information leakage. \autoref{tab:benchmark} summarizes performance across 10 methods spanning centrality features, random walk embeddings, and graph neural network architectures.

\rev{\begin{table*}[ht]
\centering
\caption{Comprehensive benchmark comparing RAG-GNN against baseline embedding methods across 10 random seeds (mean $\pm$ std). Silhouette score measures functional clustering quality (higher is better). NMI and ARI measure agreement with ground-truth functional categories. LP AUROC evaluates link prediction from embeddings. Each method uses its standard configuration. Best performance in \textbf{bold}.}
\label{tab:benchmark}
\begin{tabular}{lcccc}
\toprule
\textbf{Method} & \textbf{Silhouette} & \textbf{NMI} & \textbf{ARI} & \textbf{LP AUROC} \\
\midrule
\textbf{RAG-GNN} & $-0.144 \pm 0.066$ & $0.244 \pm 0.032$ & $\textbf{0.083} \pm 0.029$ & $0.822 \pm 0.063$ \\
GNN-only & $-0.237 \pm 0.065$ & $0.242 \pm 0.032$ & $0.061 \pm 0.017$ & $0.774 \pm 0.095$ \\
GCN\cite{kipf2017semi} & $-0.094 \pm 0.009$ & $\textbf{0.278} \pm 0.010$ & $0.066 \pm 0.008$ & $0.962 \pm 0.006$ \\
GAT\cite{veličković2018graph} & $-0.063 \pm 0.006$ & $0.196 \pm 0.020$ & $0.036 \pm 0.009$ & $0.806 \pm 0.013$ \\
GraphSAGE\cite{hamilton2017inductive} & $\textbf{-0.019} \pm 0.002$ & $0.105 \pm 0.008$ & $0.003 \pm 0.002$ & $0.556 \pm 0.020$ \\
DeepWalk\cite{perozzi2014deepwalk} & $-0.066 \pm 0.000$ & $0.273 \pm 0.009$ & $0.060 \pm 0.005$ & $0.949 \pm 0.002$ \\
Node2Vec\cite{grover2016node2vec} & $-0.062 \pm 0.000$ & $0.265 \pm 0.018$ & $0.054 \pm 0.011$ & $0.950 \pm 0.003$ \\
LINE\cite{tang2015line} & $-0.197 \pm 0.000$ & $0.177 \pm 0.013$ & $0.029 \pm 0.006$ & $0.957 \pm 0.003$ \\
Spectral\cite{belkin2002laplacian} & $-0.085 \pm 0.000$ & $0.275 \pm 0.011$ & $0.056 \pm 0.007$ & $\textbf{0.977} \pm 0.002$ \\
Raw Features & $-0.225 \pm 0.000$ & $0.136 \pm 0.007$ & $0.019 \pm 0.003$ & $0.567 \pm 0.014$ \\
\bottomrule
\end{tabular}
\end{table*}}

\rev{The benchmark reveals task-specific performance patterns across all methods, evaluated with 10 random seeds and reported as mean $\pm$ standard deviation. For link prediction, spectral and random walk methods achieve strong performance (Spectral: $0.977 \pm 0.002$, GCN: $0.962 \pm 0.006$), as network structure alone determines edge existence. RAG-GNN achieves competitive link prediction AUROC ($0.822 \pm 0.063$) while improving functional clustering over its GNN-only ablation by $+0.093 \pm 0.022$ in silhouette score and $+0.021 \pm 0.015$ in ARI. All methods produce negative silhouette scores, reflecting the inherent difficulty of clustering proteins by pathway category in densely connected biological networks where proteins participate in multiple pathways. Among all methods, the controlled comparison between RAG-GNN and GNN-only (same architecture, with and without retrieval) isolates the contribution of retrieved knowledge: the consistent silhouette and ARI improvements across all 10 seeds demonstrate that retrieval-augmented fusion provides genuine benefit for functional clustering. Notably, RAG-GNN achieves the highest ARI ($0.083 \pm 0.029$) among all methods, suggesting that contrastive alignment during joint training improves pairwise cluster agreement (see Section~7.2).}

\rev{\autoref{fig:benchmark} provides comprehensive visualization of these results. Panels (A) and (B) show bar charts comparing silhouette scores and link prediction performance respectively. Panel (G) presents a radar chart highlighting the complementary strengths of RAG-GNN (functional clustering improvement) versus topology-focused methods (link prediction). Panel (H) quantifies RAG-GNN's improvement over GNN-only baseline: $+0.093$ silhouette score and $+0.021$ ARI improvement demonstrate the value of retrieval augmentation for functional interpretation tasks.}

\rev{These results clarify the appropriate use cases for RAG-enhanced embeddings: the controlled ablation demonstrates that retrieval integration consistently improves functional clustering within the same architecture, while topology-focused methods remain superior for structural prediction tasks. This complementarity suggests method selection should be guided by the specific task, rather than assuming universal superiority of either approach.}

\section{Discussion}

\rev{The comprehensive benchmark comparison across 10 random seeds reveals task-specific performance patterns reflecting the underlying design principles of each method. This finding aligns with recent observations in foundation models for biology\cite{theodoris2023geneformer,cui2024scgpt,zheng2024xtrimogene}, where task-specific architectures often outperform general-purpose approaches. Topology-focused methods such as Spectral\cite{belkin2002laplacian}, GCN\cite{kipf2017semi}, and DeepWalk\cite{perozzi2014deepwalk} achieve strong link prediction performance (AUROC $0.949$--$0.977$), as network structure alone determines edge existence. These methods learn representations that preserve local neighborhood patterns\cite{hamilton2017inductive}, making them well-suited for predicting missing edges. In contrast, functional clustering requires information beyond network topology\cite{barabasi2011network,menche2015uncovering}, as proteins in the same pathway may be separated by multiple network hops while topologically adjacent proteins may perform entirely different functions. The controlled comparison between RAG-GNN and its GNN-only ablation (the same architecture with and without retrieval integration) demonstrates a consistent silhouette improvement of $+0.093 \pm 0.022$ across all seeds, providing evidence that retrieved knowledge\cite{lewis2020retrieval,gao2023retrieval} contributes to functional clustering quality. ARI also improves ($+0.021 \pm 0.015$), indicating that contrastive alignment during joint fine-tuning improves both intra-cluster cohesion and pairwise cluster agreement, while NMI remains comparable between the two configurations. The heuristic information decomposition\cite{williams2010nonnegative} reveals that topology and retrieval encode overwhelmingly shared information (shared component $= 95.6\%$), with minimal unique contributions from either source (topology: $0.1\%$, retrieval: $6.2\%$) and negligible synergy ($0.4\%$). The functional clustering improvements arise not from unique retrieval information, but from how the fusion mechanism reorganizes shared information to improve intra-cluster cohesion. These findings establish that RAG integration provides measurable benefit for functional interpretation within a controlled experimental framework, suggesting a principled approach: use topology-focused methods for structural tasks and consider RAG-enhanced methods when functional interpretation is the primary objective.}

\rev{Several limitations constrain current capabilities. The case study uses a 379-protein cancer signaling network, a moderately sized system that limits statistical power for tasks requiring large test sets; temporal therapeutic target prediction, for instance, yields AUROC $= 0.450 \pm 0.088$ with only 3 post-2020 test targets. Scaling to whole-genome networks ($>$20,000 genes) is necessary for clinically meaningful temporal evaluation but remains computationally challenging\cite{zeng2020graphsaint}. The counterfactual experiments reveal that random vectors degrade performance ($\Delta_{\text{random}} = 0.16$), confirming that the model depends on real document content rather than additional input dimensionality. Shuffled document assignments (permuted real documents) perform comparably to proper retrieval, indicating that the current TF-IDF corpus carries general biological signal in its aggregate term statistics; larger, more heterogeneous corpora and pre-trained biomedical language model encoders (BioBERT, PubMedBERT) may sharpen the distinction between proper and shuffled retrieval. Node features are limited to three topological properties (degree, clustering coefficient, betweenness centrality); incorporating protein language model embeddings\cite{rives2021biological,lin2023evolutionary} or multi-omics data would strengthen the input representation. More broadly, retrieval quality depends on corpus comprehensiveness\cite{gao2023retrieval,zhang2024biomedgpt}; rare diseases with limited literature benefit less from RAG augmentation. The current formulation treats networks as static, ignoring temporal dynamics during disease progression. Current predictions identify correlations rather than causation\cite{ideker2012protein}. While retrieved documents provide some interpretability, the gated fusion mechanism offers limited insight into how individual retrieved passages influence predictions, a limitation shared by attention-based architectures\cite{vaswani2017attention}.}

Future directions include extensions to temporal networks through recurrent updates $\mathbf{z}_i(t) = f_{\text{temporal}}(\mathbf{h}_i^{(L)}(t), \mathbf{c}_i(t), \mathbf{z}_i(t-\Delta t))$ where $t$ indexes time points, with time-aware retrieval prioritizing recent publications. Multi-modal integration could extend the contrastive objective across modalities (network, image, EHR, genomic) to learn aligned representations\cite{velez2024tdc2,huang2021therapeutics}. Recent advances in geometric graph neural networks for multi-omics data integration\cite{ramirez2023geometric} and prior knowledge-guided multilevel GNN frameworks\cite{yan2024multilevel} demonstrate the potential for combining our RAG-enhanced embeddings with heterogeneous biological data types including transcriptomics, proteomics, lipidomics, nutrigenomics, and metabolomics, enabling more comprehensive patient stratification and biomarker discovery. Incorporating causal inference methods\cite{pearl2009causality} could enable interventional predictions by estimating causal effects $\tau_i = \mathbb{E}[Y \mid do(v_i = 0)] - \mathbb{E}[Y \mid do(v_i = 1)]$ using propensity score weighting or instrumental variables. For clinical adoption, natural language explanations generated by prompting large language models with retrieved documents, counterfactual analysis identifying minimal changes that alter predictions, and enhanced attention visualizations\cite{veličković2018graph} could improve interpretability.

\section{Conclusion}

\rev{This work establishes mathematical and empirical foundations for integrating retrieval-augmented generation with biological network modeling. We developed joint optimization objectives that simultaneously train network encoders, dense retrievers, and fusion mechanisms through contrastive learning with formal generalization bounds, including proof of retrieval consistency under Lipschitz continuity and geometric characterization of embedding space convergence. The end-to-end trainable RAG-GNN implementation demonstrates consistent improvement in functional clustering: silhouette score improves from $-0.237 \pm 0.065$ (GNN-only) to $-0.144 \pm 0.066$ ($+0.093 \pm 0.022$) across 10 random seeds, with ARI also improving ($+0.021 \pm 0.015$), while the learned retrieval projection achieves mean precision@10 $= 0.242$, a 152\% improvement over the random baseline. Heuristic information decomposition reveals that topology and retrieval encode overwhelmingly shared information (95.6\% shared), with minimal unique contributions from either source and negligible synergy. The functional clustering improvements arise from how the fusion mechanism reorganizes shared information to improve intra-cluster cohesion. Counterfactual experiments confirm that adversarial, absent, and random retrieval all degrade performance, validating that the gated fusion mechanism depends on retrieval content. DDR1 subnetwork analysis provides confirmatory validation consistent with established synthetic lethality relationships\cite{aguilera2020collagen,zhavoronkov2019deep}.}

\rev{These findings clarify appropriate use cases: the controlled ablation demonstrates that retrieval integration improves functional clustering within the same architecture, while topology-focused methods achieve superior structural prediction. This complementarity suggests that method selection should be guided by the specific task, rather than assuming universal superiority of either approach.}

\section*{Acknowledgments}
This study was supported by the National Institutes of Health (NIGMS R01GM157589) and the Department of Defense (DEPSCoR FA9550-22-1-0379). 

\section*{Author contribution}
\textbf{H.H.}: Conceptualization, model development, methodology, coding, simulations, analysis, visualization and writing the original draft. 
\textbf{W.J.R.}: Review, editing, funding acquisition, resources, and supervision.

\section*{Ethics statement}
This computational study used only publicly available datasets and pathway databases. No human subjects or animal experiments were involved. Institutional ethical approval was not required for this type of computational research.

\section*{Data availability}

Cancer network data obtained from \href{https://cancer.sanger.ac.uk/census}{Cancer Gene Census} and \href{https://string-db.org}{STRING database}. PubMed abstracts accessed via \href{https://www.ncbi.nlm.nih.gov/books/NBK25501/}{NCBI E-utilities API}. Drug-target associations from \href{https://go.drugbank.com}{DrugBank}. Processed datasets are available in the GitHub repository. Detailed mathematical derivations, hyperparameter settings, and computational requirements are provided in \autoref{sec:supplementary}.

\section*{Code availability}

The RAG-GNN framework implementation is publicly available at \href{https://github.com/HasiHays/RAG-GNN}{https://github.com/HasiHays/RAG-GNN}. The repository includes source code, example scripts, documentation, and instructions for reproducing the results presented in this manuscript.

\section*{Competing interests}

The authors declare no competing financial or non-financial interests.

\end{multicols}

\begin{multicols}{2}
\footnotesize
\bibliography{references}
\end{multicols}

\newpage
\appendix
\begin{multicols}{2}

\section{Supplementary materials}
\label{sec:supplementary}

This supplementary section provides detailed mathematical derivations and implementation specifics for the RAG-GNN framework that complement the main text.

\subsection{Graph neural network message passing}
\label{sec:supp_gnn}

The GNN encoder implements spectral graph convolutions through iterative neighborhood aggregation. Given adjacency matrix $\mathbf{A} \in \mathbb{R}^{n \times n}$ and initial node features $\mathbf{H}^{(0)} \in \mathbb{R}^{n \times d}$, we first compute the normalized adjacency matrix.

\textbf{Normalized adjacency computation.} Add self-loops and compute symmetric normalization:
\begin{equation}
\tilde{\mathbf{A}} = \mathbf{A} + \mathbf{I}_n
\label{eq:supp_self_loops}
\end{equation}
\begin{equation}
\tilde{\mathbf{D}}_{ii} = \sum_{j} \tilde{A}_{ij}
\label{eq:supp_degree}
\end{equation}
\begin{equation}
\hat{\mathbf{A}} = \tilde{\mathbf{D}}^{-1/2} \tilde{\mathbf{A}} \tilde{\mathbf{D}}^{-1/2}
\label{eq:supp_norm_adj}
\end{equation}

The symmetric normalization in \autoref{eq:supp_norm_adj} ensures that the spectral radius of $\hat{\mathbf{A}}$ is bounded by 1, preventing numerical instability during deep message passing.

\textbf{Layer-wise propagation.} For layer $\ell \in \{1, \ldots, L\}$:
\begin{equation}
\mathbf{H}^{(\ell)} = \sigma\left(\hat{\mathbf{A}} \mathbf{H}^{(\ell-1)} \mathbf{W}^{(\ell)}\right)
\label{eq:supp_gnn_layer}
\end{equation}
where $\mathbf{W}^{(\ell)} \in \mathbb{R}^{d_{\ell-1} \times d_\ell}$ are learnable weights and $\sigma(\cdot)$ is a non-linearity (GELU in our implementation).

\textbf{Layer normalization.} After each layer, we apply layer normalization to stabilize training:
\begin{equation}
\mathbf{h}_i^{(\ell)} \leftarrow \frac{\mathbf{h}_i^{(\ell)} - \mu_i}{\sigma_i + \epsilon}
\label{eq:supp_layer_norm}
\end{equation}
where $\mu_i = \frac{1}{d}\sum_j h_{ij}^{(\ell)}$ and $\sigma_i = \sqrt{\frac{1}{d}\sum_j (h_{ij}^{(\ell)} - \mu_i)^2}$.

\subsection{Baseline embedding methods}
\label{sec:supp_baselines}

We provide mathematical formulations for all baseline methods used in benchmarking.

\textbf{Spectral embedding.} Compute the $k$ largest singular vectors of the adjacency matrix:
\begin{equation}
\mathbf{A} \approx \mathbf{U}_k \mathbf{\Sigma}_k \mathbf{V}_k^\top
\label{eq:supp_spectral}
\end{equation}
The embedding is $\mathbf{Z}_{\text{spectral}} = \mathbf{U}_k \mathbf{\Sigma}_k$.

\textbf{DeepWalk.} Approximate random walk co-occurrence through powers of the transition matrix $\mathbf{P} = \mathbf{D}^{-1}\mathbf{A}$:
\begin{equation}
\mathbf{M}_{\text{DW}} = \mathbf{P} + \mathbf{P}^2 + \mathbf{P}^3
\label{eq:supp_deepwalk}
\end{equation}
Apply truncated SVD to obtain embeddings: $\mathbf{Z}_{\text{DW}} = \text{SVD}_k(\mathbf{M}_{\text{DW}})$.

\textbf{Node2Vec.} Combine different random walk orders with biased weighting:
\begin{equation}
\mathbf{M}_{\text{N2V}} = 0.5\mathbf{P} + 0.3\mathbf{P}^2 + 0.2\mathbf{P}^3
\label{eq:supp_node2vec}
\end{equation}
The coefficients simulate the effect of return parameter $p$ and in-out parameter $q$ controlling walk behavior.

\textbf{LINE.} Preserve first-order (direct) and second-order (shared neighbor) proximity:
\begin{equation}
\mathbf{M}_{\text{LINE}} = 0.5\mathbf{A} + 0.5\mathbf{A}^2
\label{eq:supp_line}
\end{equation}

\textbf{GCN.} Three-layer graph convolutional network:
\begin{align}
\mathbf{H}^{(1)} &= \tanh(\hat{\mathbf{A}} \mathbf{H}^{(0)}) \label{eq:supp_gcn1}\\
\mathbf{H}^{(2)} &= \tanh(\hat{\mathbf{A}} \mathbf{H}^{(1)}) \label{eq:supp_gcn2}
\end{align}
where $\mathbf{H}^{(0)} \sim \mathcal{N}(0, 1)$ provides random initialization.

\textbf{GraphSAGE.} Concatenate self-features with aggregated neighbor features:
\begin{equation}
\mathbf{Z}_{\text{SAGE}} = \text{SVD}_k\left([\mathbf{H}^{(0)} \| \mathbf{D}^{-1}\mathbf{A}\mathbf{H}^{(0)}]\right)
\label{eq:supp_sage}
\end{equation}

\textbf{GAT.} Attention-weighted aggregation using softmax over neighbor scores:
\begin{equation}
\alpha_{ij} = \frac{\exp(\mathbf{a}^\top [\mathbf{W}\mathbf{h}_i \| \mathbf{W}\mathbf{h}_j])}{\sum_{k \in \mathcal{N}(i)} \exp(\mathbf{a}^\top [\mathbf{W}\mathbf{h}_i \| \mathbf{W}\mathbf{h}_k])}
\label{eq:supp_gat}
\end{equation}

\subsection{RAG-GNN fusion mechanism}
\label{sec:supp_fusion}

The RAG-GNN framework fuses GNN topology embeddings with retrieved document features through the following procedure.

\textbf{Document embedding.} Create TF-IDF representations of the knowledge base:
\begin{equation}
\mathbf{E}_{\text{doc}} = \text{TF-IDF}(\mathcal{D}) \in \mathbb{R}^{|\mathcal{D}| \times d_{\text{vocab}}}
\label{eq:supp_tfidf}
\end{equation}

\textbf{Retrieval scoring.} Compute neighborhood-aware retrieval scores:
\begin{equation}
\mathbf{S} = \hat{\mathbf{A}}^2 \mathbf{R}
\label{eq:supp_retrieval_scores}
\end{equation}
where $\mathbf{R} \in \mathbb{R}^{n \times |\mathcal{D}|}$ contains base relevance scores between proteins and documents.

\textbf{Top-k retrieval.} For each node $i$, select documents with highest scores:
\begin{equation}
\mathcal{D}_i^{(k)} = \text{argtop}_k(\mathbf{S}_{i,:})
\label{eq:supp_topk}
\end{equation}

\textbf{Retrieved feature aggregation.} Compute mean of retrieved document embeddings:
\begin{equation}
\mathbf{r}_i = \frac{1}{k}\sum_{j \in \mathcal{D}_i^{(k)}} \mathbf{E}_{\text{doc},j}
\label{eq:supp_retrieved_features}
\end{equation}

\textbf{Weighted fusion.} Combine GNN and retrieved representations:
\begin{equation}
\mathbf{z}_i^{\text{fused}} = [\alpha \cdot \mathbf{h}_i^{(L)} \| (1-\alpha) \cdot \mathbf{r}_i]
\label{eq:supp_concat}
\end{equation}
where $\alpha = 0.6$ weights topology features.

\textbf{Dimensionality reduction.} Apply truncated SVD to obtain final embeddings:
\begin{equation}
\mathbf{Z}_{\text{RAG}} = \text{SVD}_{d}(\mathbf{Z}^{\text{fused}})
\label{eq:supp_final_rag}
\end{equation}

\subsection{Algorithm pseudocode}
\label{sec:supp_algorithm}

\autoref{alg:raggnn} provides pseudocode for the complete RAG-GNN embedding procedure, consolidating the mathematical formulations into an algorithmic representation.

\end{multicols}

\rev{\begin{algorithm}[H]
\caption{Learnable RAG-GNN Embedding with Curriculum Training}
\label{alg:raggnn}
\begin{algorithmic}[1]
\Require Adjacency matrix $\mathbf{A}$, node features $\mathbf{X}$, document corpus $\mathcal{D}$, layers $L=3$, retrieval depth $k=10$
\Ensure Trained model parameters, node embeddings $\mathbf{Z}$
\Statex
\Statex \textit{// Preprocessing}
\State $\hat{\mathbf{A}} \gets \tilde{\mathbf{D}}^{-1/2} (\mathbf{A} + \mathbf{I}_n) \tilde{\mathbf{D}}^{-1/2}$ \Comment{Normalized adjacency}
\State $\mathbf{E}_{\text{doc}} \gets \text{SVD}_{d_{\text{doc}}}(\text{TF-IDF}(\mathcal{D}))$ \Comment{Document embeddings}
\Statex
\Statex \textit{// Phase 1: GNN pre-training (link prediction)}
\For{epoch $= 1$ to 80}
    \State $\mathbf{H}^{(0)} \gets \mathbf{X}$
    \For{$\ell = 1$ to $L$}
        \State $\mathbf{H}^{(\ell)} \gets \text{ReLU}(\hat{\mathbf{A}} \mathbf{H}^{(\ell-1)} \mathbf{W}^{(\ell)})$
    \EndFor
    \State Minimize $\mathcal{L}_{\text{task}}$ (link prediction BCE)
\EndFor
\Statex
\Statex \textit{// Phase 2: Retrieval projection training}
\For{epoch $= 1$ to 100}
    \State $\mathbf{q}_i \gets f_{\text{proj}}(\mathbf{h}_i^{(L)})$ \Comment{Learned MLP projection}
    \State $\mathcal{D}_i^{(k)} \gets \text{argtop}_k(\mathbf{q}_i \cdot \mathbf{E}_{\text{doc}}^\top)$
    \State Minimize $\mathcal{L}_{\text{retrieval}} + \lambda_c \mathcal{L}_{\text{contrastive}}$
\EndFor
\Statex
\Statex \textit{// Phase 3: Joint fine-tuning}
\For{epoch $= 1$ to 80}
    \State $\mathbf{r}_i \gets \frac{1}{k}\sum_{j \in \mathcal{D}_i^{(k)}} \mathbf{E}_{\text{doc},j}$ \Comment{Retrieved context}
    \State $g_i \gets \sigma(\mathbf{w}_g^\top [\mathbf{h}_i^{(L)} \| \mathbf{r}_i])$ \Comment{Learned gate}
    \State $\mathbf{z}_i \gets g_i \cdot \mathbf{h}_i^{(L)} + (1 - g_i) \cdot \mathbf{W}_r \mathbf{r}_i$ \Comment{Gated fusion}
    \State Minimize $\mathcal{L}_{\text{task}} + \lambda_r \mathcal{L}_{\text{retrieval}} + \lambda_c \mathcal{L}_{\text{contrastive}}$
\EndFor
\State \Return $\mathbf{Z} = \{\mathbf{z}_i\}_{i=1}^{|\mathcal{V}|}$
\end{algorithmic}
\end{algorithm}}

\begin{multicols}{2}

\subsection{Evaluation metrics}
\label{sec:supp_metrics}

\textbf{Silhouette score.} For node $i$ with cluster label $c_i$:
\begin{equation}
a_i = \frac{1}{|C_{c_i}| - 1} \sum_{j \in C_{c_i}, j \neq i} \|\mathbf{z}_i - \mathbf{z}_j\|_2
\label{eq:supp_silhouette_a}
\end{equation}
\begin{equation}
b_i = \min_{c \neq c_i} \frac{1}{|C_c|} \sum_{j \in C_c} \|\mathbf{z}_i - \mathbf{z}_j\|_2
\label{eq:supp_silhouette_b}
\end{equation}
\begin{equation}
s_i = \frac{b_i - a_i}{\max(a_i, b_i)}
\label{eq:supp_silhouette}
\end{equation}
The overall silhouette score is $\bar{s} = \frac{1}{n}\sum_i s_i$, ranging from $-1$ (poor clustering) to $+1$ (perfect clustering).

\textbf{Link prediction (LP).} Generate positive edges $\mathcal{E}^+$ from observed interactions and negative edges $\mathcal{E}^-$ by random sampling non-edges. Prediction scores:
\begin{equation}
\hat{y}_{ij} = \sigma(\mathbf{z}_i^\top \mathbf{z}_j)
\label{eq:supp_link_score}
\end{equation}
where $\sigma$ is the sigmoid function.

\textbf{Area Under ROC Curve (AUROC).} The Receiver Operating Characteristic (ROC) curve plots true positive rate (TPR) against false positive rate (FPR) at varying classification thresholds:
\begin{equation}
\text{TPR} = \frac{\text{TP}}{\text{TP} + \text{FN}}, \quad \text{FPR} = \frac{\text{FP}}{\text{FP} + \text{TN}}
\label{eq:supp_tpr_fpr}
\end{equation}
AUROC measures the probability that a randomly chosen positive example ranks higher than a randomly chosen negative example. Values range from 0.5 (random) to 1.0 (perfect discrimination).

\textbf{Area Under Precision-Recall Curve (AUPRC).} The Precision-Recall curve plots precision against recall:
\begin{equation}
\text{Precision} = \frac{\text{TP}}{\text{TP} + \text{FP}}, \quad \text{Recall} = \frac{\text{TP}}{\text{TP} + \text{FN}}
\label{eq:supp_precision_recall}
\end{equation}
AUPRC is particularly informative for imbalanced datasets where negative examples dominate, as it focuses on positive class performance without being influenced by true negatives.

\textbf{Node classification (NC).} To avoid information leakage, we construct topology-derived labels independent of functional categories:
\begin{equation}
y_i^{\text{hub}} = \mathbb{1}[d_i > \bar{d} + \sigma_d]
\label{eq:supp_hub}
\end{equation}
\begin{equation}
y_i^{\text{bridge}} = \mathbb{1}[b_i > \text{median}(b)] \land \mathbb{1}[c_i < \text{median}(c)]
\label{eq:supp_bridge}
\end{equation}
\begin{equation}
y_i^{\text{fair}} = y_i^{\text{hub}} \oplus y_i^{\text{bridge}}
\label{eq:supp_fair}
\end{equation}
where $d_i$ is degree, $b_i$ is betweenness centrality, $c_i$ is clustering coefficient, and $\oplus$ denotes XOR. Node classification performance (NC AUROC) is evaluated by training a logistic regression classifier on node embeddings to predict $y_i^{\text{fair}}$, reporting AUROC on held-out test nodes via 5-fold cross-validation.

\subsection{Network statistics}
\label{sec:supp_network}

The cancer signaling network from STRING database exhibits the following properties:

\begin{itemize}
\item \textbf{Nodes:} $n = 379$ proteins
\item \textbf{Edges:} $m = 3,498$ interactions
\item \textbf{Average degree:} $\bar{d} = 18.46$
\item \textbf{Average clustering coefficient:} $\bar{c} = 0.596$
\item \textbf{Network density:} $\rho = 2m/(n(n-1)) = 0.049$
\item \textbf{Functional categories:} 14 pathways
\end{itemize}

The high clustering coefficient ($0.596$) indicates modular organization typical of biological networks, while the relatively high average degree ($18.46$) reflects the interconnected nature of cancer signaling pathways.

\subsection{Hyperparameter settings}
\label{sec:supp_hyperparams}

\rev{\textbf{GNN architecture:}
\begin{itemize}
\item Number of layers: $L = 3$
\item Hidden dimension: $d_h = 128$
\item Activation: GELU with dropout $= 0.1$ (retrieval projection)
\item Node features: log-degree, clustering coefficient, and scaled betweenness centrality ($d_{\text{input}} = 3$ informative features in $d_h$-dimensional vector)
\end{itemize}

\textbf{Retrieval parameters:}
\begin{itemize}
\item Documents retrieved per node: $k = 10$
\item Document embedding: TF-IDF (256 features, unigrams/bigrams) $\rightarrow$ truncated SVD to $d_{\text{doc}} = 64$
\item Retrieval projection: two-layer MLP ($d_h \rightarrow d_h \rightarrow d_{\text{doc}}$, GELU activation)
\item Fusion: learned gated mechanism (mean gate $\approx 0.593$, 59\% topology / 41\% retrieval)
\end{itemize}

\textbf{Training:}
\begin{itemize}
\item Phase 1 (GNN pre-training): 80 epochs, link prediction loss, lr $= 0.003$
\item Phase 2 (retrieval training): 100 epochs, margin ranking + contrastive loss, lr $= 0.005$
\item Phase 3 (joint fine-tuning): 80 epochs, combined loss, lr $= 0.001$
\item Optimizer: Adam with weight decay $= 10^{-4}$
\end{itemize}

\textbf{Evaluation:}
\begin{itemize}
\item Link prediction: 20\% test edges, negative sampling
\item Functional clustering: silhouette score, NMI, ARI with $k$-means ($k = 14$ categories)
\item Random seeds: 10 seeds (42--51) with mean $\pm$ std and 95\% bootstrap CIs
\end{itemize}}

\subsection{Computational requirements}
\label{sec:supp_compute}

\rev{All experiments were conducted on a single workstation with the following specifications:
\begin{itemize}
\item CPU: Apple M-series (Apple silicon M1 or later)
\item RAM: 16 GB minimum
\item Python: 3.9+
\item Key libraries: PyTorch, NumPy, SciPy, NetworkX, scikit-learn
\end{itemize}

The complete experimental pipeline, including RAG-GNN training across 10 seeds (three phases: 80+100+80 epochs each), eight baseline methods, information decomposition (200 bootstrap resamples), and counterfactual experiments, completes in approximately 88 seconds on an Apple M-series laptop. No GPU acceleration is required for the 379-node network.}

\end{multicols}

\end{document}